\documentclass[aps,prd,onecolumn,showpacs,showkeys,amsmath,amssymb]{revtex4}
\usepackage{amsfonts}
\usepackage{graphicx}
\usepackage{dcolumn}
\usepackage{bm}
\usepackage[dvips]{color}
\begin{document}
%
%
\title{The Possible $J^{PC} = 0^{--}$ Exotic State}
\author{Chun-Kun Jiao}
\email{ckjiao@163.com}
\author{Wei Chen}
\email{boya@pku.edu.cn}
\author{Hua-Xing Chen}
\email{chx@water.pku.edu.cn}
\author{Shi-Lin Zhu}
\email{zhusl@pku.edu.cn} \affiliation{Department of Physics
and State Key Laboratory of Nuclear Physics and Technology\\
Peking University, Beijing 100871, China  }
\begin{abstract}

In order to explore the possible existence of the exotic $0^{--}$
state, we have constructed the tetraquark interpolating operators
systematically. As a byproduct, we notice the $0^{+-}$ tetraquark
operators without derivatives do not exist. The special Lorentz
structure of the $0^{--}$ currents forbids the four-quark type of
corrections to the spectral density. Now the gluon condensates are
the dominant power corrections. None of the seven interpolating
currents supports a resonant signal. Therefore we conclude that
the exotic $0^{--}$ state does not exist below 2 GeV, which is
consistent with the current experimental observations.

\end{abstract}

\keywords{exotic state, QCD sum rule}

\pacs{12.39.Mk, 11.40.-q, 12.38.Lg}
\maketitle
\pagenumbering{arabic}
%
%
\section{Introduction}\label{sec:intro}
%

Most of the experimentally observed hadrons can be interpreted as $q
\bar q$/$qqq$ states and accommodated in the quark
model~\cite{Amsler:2008zzb,Klempt:2007cp}. Up to now there has
accumulated some evidence of the exotic state with $J^{PC} = 1^{-+}$
~\cite{Adams:2006sa,Abele:1999tf,Thompson:1997bs}. Such a quantum
number is not accessible for a pair of quark and anti-quark. It is
sometimes labelled as an exotic hybrid meson with the particle
contents $\bar q g_s G_{\mu\nu} \gamma^\nu q$. Recently we have
investigated the $1^{-+}$ state using the tetraquark currents
~\cite{Chen:2008qw}. The extracted mass and characteristic decay
pattern are quite similar to those expected for the exotic hybrid
meson. Such a result is expected. Since the gluon field creates a
pair of $q\bar q$ easily, the hybrid operator $\bar q g_s G_{\mu\nu}
\gamma^\nu q$ transforms into a tetraquark interpolating operator
with the same exotic quantum number. In quantum field theory
different operators with the same quantum number mix and tend to
couple to the same physical state.

Using the same tetraquark formalism developed in the study of the
low-lying scalar mesons ~\cite{Chen:2006hy} and the exotic
$1^{-+}$ mesons ~\cite{Chen:2008qw}, we study the possible $J^{PC}
=0^{--}$ states composed of light quarks. For a neutral quark
model state $q\bar q$, we know that $J=0$ ensures $L=S$ hence
$C=(-)^{L+S}=+1$. In other words, states with $J^{PC} =0^{--},
0^{+-}$ are strictly forbidden. On the other hand, the gauge
invariant scalar and pseudoscalar operators composed of a pair of
the gluon field are $g^2_sG_{\mu\nu}^a G^{a\mu\nu}$ and
$\epsilon^{\mu\nu\alpha\beta}g^2_sG_{\mu\nu}^a
G^{a}_{\alpha\beta}$, both of which carry the even C-parity.

We construct all the local tetraquark currents with $J^{PC} =
0^{--}$. There are two kinds of constructions: $(qq)(\bar q \bar
q)$ and $(\bar q q)(\bar q q)$. They can be related to each other
by using the Fierz transformation. As usual, we use the first
set~\cite{Chen:2006hy}. Their flavor structure can be
$\mathbf{\bar 3}_f \otimes \mathbf{3}_f$, $\mathbf{6}_f \otimes
\mathbf{\bar 6}_f$, and $\mathbf{\bar 3}_f \otimes \mathbf{\bar
6}_f \oplus \mathbf{6}_f \otimes \mathbf{3}_f$ ($(qq)(\bar q \bar
q)$). With all these independent currents, we perform the QCD sum
rule analysis. As a byproduct, we notice that there does not exist
any tetraquark interpolating operator without derivative for the
$J^{PC} = 0^{+-}$ case.

This paper is organized as follows. In Sec.~\ref{sec:current}, we
construct the tetraquark currents with $J^{PC} = 0^{--}$ using the
diquark ($qq$) and antidiquark ($\bar q \bar q$) fields. The
tetraquark currents constructed with the quark-antiquark ($\bar q
q$) pairs are shown in Appendix.\ref{app}. We present the spectral
density in Sec.~\ref{sec:ope} and perform the numerical analysis
in Sec.~\ref{sec:num}. For comparison, we present the finite
energy sum rule analysis in the Appendix.\ref{sec:FESR}. The last
section is a short summary.

%
%
\section{tetraquark interpolating currents}\label{sec:current}
\subsection{The $J^{PC}=0^{--}$ Tetraquark Interpolating Currents}
%
%
In this section, we construct the tetraquark interpolating currents
with $J^{PC}=0^{--}$ using diquark and antidiquark fields. Such a
quantum number can not be accessed by a $q\bar q$ pair. The currents
can be similarly constructed by using the quark-antiquark pairs.
However, as shown in Appendix.\ref{app}, these two constructions are
equivalent as we have shown several times in our previous
studies~\cite{Chen:2006hy,Chen:2008qw}.

The pseudoscalar tetraquark currents can be constructed using five
independent diquark fields, which are constructed by five
independent $\gamma$-matrices
%
\begin{eqnarray}
\nonumber &&
S_{abcd}=(q_{1a}^TCq_{2b})(\bar{q}_{3c}\gamma_5C\bar{q}^T_{4d}) \, ,\\
\nonumber &&
V_{abcd}=(q_{1a}^TC\gamma_5q_{2b})(\bar{q}_{3c}C\bar{q}^T_{4d}) \, ,\\
&& T_{abcd}=(q_{1a}^TC\sigma_{\mu\nu}q_{2b})(\bar{q}_{3c}\sigma^{\mu\nu}\gamma_5C\bar{q}^T_{4d}) \, ,\\
\nonumber &&
A_{abcd}=(q_{1a}^TC\gamma_{\mu}q_{2b})(\bar{q}_{3c}\gamma^{\mu}\gamma_5C\bar{q}^T_{4d}) \, ,\\
\nonumber &&
P_{abcd}=(q_{1a}^TC\gamma_{\mu}\gamma_5q_{2b})(\bar{q}_{3c}\gamma^{\mu}C\bar{q}^T_{4d})
.
\end{eqnarray}
%
where $q_{1-4}$ represents the $up$, $down$ and $strange$ quarks,
and $a-d$ are the color indices.

To compose a color singlet pseudoscalar tetraquark current, the
diquark and antidiquark should have the same color and spin
symmetries. So the color structure of the tetraquark is either
$\mathbf 6 \otimes \mathbf { \bar 6}$ or $\mathbf { \bar 3 }\otimes
\mathbf 3$, which is denoted by labels  $\mathbf  6 $ and $\mathbf
3 $ respectively. Therefore, considering both the color and Lorentz
structures, there are altogether ten terms of products
\begin{eqnarray}
\{S \oplus V \oplus T \oplus A \oplus P\}_{Lorentz} \otimes \{3
\oplus 6\}_{Color}.
\end{eqnarray}
We list them as follows
%
\begin{eqnarray}\label{currents}
\nonumber  6_F \otimes \bar{6}_F ~ (S) &&\left \{
\begin{array}{l}
S_6 =
q_{1a}^TCq_{2b}(\bar{q}_{3a}\gamma_5C\bar{q}^T_{4b}+\bar{q}_{3b}\gamma_5C\bar{q}^T_{4a})
\, , \\
V_6=q_{1a}^TC\gamma_5q_{2b}(\bar{q}_{3a}C\bar{q}^T_{4b}+\bar{q}_{3b}C\bar{q}^T_{4a})
\, , \\
T_3=q_{1a}^TC\sigma_{\mu\nu}q_{2b}(\bar{q}_{3a}\sigma^{\mu\nu}\gamma_5C\bar{q}^T_{4b}-\bar{q}_{3b}\sigma^{\mu\nu}
\gamma_5C\bar{q}^T_{4a}) \, ,
\end{array} \right.
\\
\bar{3}_F \otimes 3_F ~ (A) && \left \{
\begin{array}{l}
S_3=q_{1a}^TCq_{2b}(\bar{q}_{3a}\gamma_5C\bar{q}^T_{4b}-\bar{q}_{3b}\gamma_5C\bar{q}^T_{4a})
\, ,
\\
V_3=q_{1a}^TC\gamma_5q_{2b}(\bar{q}_{3a}C\bar{q}^T_{4b}-\bar{q}_{3b}C\bar{q}^T_{4a})
\, , \\
T_6=q_{1a}^TC\sigma_{\mu\nu}q_{2b}(\bar{q}_{3a}\sigma^{\mu\nu}\gamma_5C\bar{q}^T_{4b}+\bar{q}_{3b}\sigma^{\mu\nu}
\gamma_5C\bar{q}^T_{4a}) \, ,
\end{array} \right.
\\ \nonumber
\bar{3}_F \otimes \bar{6}_F ~ (M) && \left \{
\begin{array}{l}
A_6=q_{1a}^TC\gamma_{\mu}q_{2b}(\bar{q}_{3a}\gamma^{\mu}\gamma_5C\bar{q}^T_{4b}+\bar{q}_{3b}\gamma^{\mu}
\gamma_5C\bar{q}^T_{4a}) \, , \\
P_3=q_{1a}^TC\gamma_{\mu}\gamma_5q_{2b}(\bar{q}_{3a}\gamma^{\mu}C\bar{q}^T_{4b}-\bar{q}_{3b}\gamma^{\mu}C\bar{q}^T_{4a})
\, .
\end{array} \right.
\\ \nonumber 6_F \otimes 3_F ~ (M) && \left \{
\begin{array}{l}
P_6=q_{1a}^TC\gamma_{\mu}\gamma_5q_{2b}(\bar{q}_{3a}\gamma^{\mu}C\bar{q}^T_{4b}+\bar{q}_{3b}\gamma^{\mu}C\bar{q}^T_{4a})
\, , \\
A_3=q_{1a}^TC\gamma_{\mu}q_{2b}(\bar{q}_{3a}\gamma^{\mu}\gamma_5C\bar{q}^T_{4b}-\bar{q}_{3b}\gamma^{\mu}
\gamma_5C\bar{q}^T_{4a}) \, .
\end{array} \right.
\end{eqnarray}
%
In the above expressions, the flavor structure is fixed at the same
time due to the Pauli principle. The currents $S_6$, $V_6$, $T_3$
belong to the symmetric flavor representation $\mathbf {6_F }\otimes
\mathbf {\bar{6}_F }(S)$ where both diquark and antidiquark fields
have the symmetric flavor structure. The currents $S_3$, $V_3$,
$T_6$ belong to the antisymmetric flavor representation $\mathbf
{\bar{3}_F} \otimes \mathbf { 3_F }(A)$, where both diquark and
antidiquark fields have the antisymmetric flavor structure. $A_6$,
$P_3$ for $\mathbf {\bar{3}_F }\otimes \mathbf {\bar{6}_F} (M)$ and
$A_3$, $P_6$ for $\mathbf {6_F }\otimes \mathbf {3_F }(M)$, where
$M$ represents the mixed flavor symmetry. The isovector states with
charges can be observed in the experiments more easily, therefore in
this paper we concentrate on the isovector currents which was shown
in the $SU(3)$ tetraquark weight diagram in
Fig~1~\cite{Chen:2008qw}. We have:
\begin{eqnarray}\label{content}
\nonumber qq\bar q\bar q (S),~qs\bar q\bar s (S)~~ &\in& ~~ 6_F
\otimes \bar{6}_F ~~~ (S) \, , \\
 qs\bar q\bar s (A)~~ &\in& ~~ \bar{3}_F
\otimes 3_F ~~~ (A) \, , \\
\nonumber  qq\bar q\bar q (M), qs\bar q\bar s (M)~~ &\in& ~~
(\bar{3}_F \otimes \bar{6}_F) \oplus (6_F \otimes 3_F) ~~~ (M) \, .
\end{eqnarray}
We do not differentiate $up$ and $down$ quarks and denote them by
$q$. We are only interested in those neutral components. The other
states do not carry definite C-parity. It turns out that the
neutral isovector and isoscalar states have the same QCD sum
rules. Our following discussions are valid for both of them.
Taking the charge-conjugation transformation, we get
\begin{equation}
\mathbb{C}S_6\mathbb{C}^{-1}=V_6 \, ,
\mathbb{C}A_6\mathbb{C}^{-1}=P_6 \, ,
\mathbb{C}A_3\mathbb{C}^{-1}=P_3 \, ,
\mathbb{C}S_3\mathbb{C}^{-1}=V_3 \,
,\mathbb{C}T_6\mathbb{C}^{-1}=T_6 \, ,
\mathbb{C}T_3\mathbb{C}^{-1}=T_3 \, .
\end{equation}
$T_6$ and $T_3$ have even charge-conjugation parity. We conclude
that the currents with $J^{PC}=0^{--}$ are:
\begin{eqnarray}
\nonumber \eta^{(S)} &=&
S_6-V_6=q_{1a}^TCq_{2b}(\bar{q}_{3a}\gamma_5C\bar{q}^T_{4b}+\bar{q}_{3b}\gamma_5C\bar{q}^T_{4a})
-q_{1a}^TC\gamma_5q_{2b}(\bar{q}_{3a}C\bar{q}^T_{4b}+\bar{q}_{3b}C\bar{q}^T_{4a})\,
,
\\ \nonumber
\eta^{(M)}_{1} &=&
A_6-P_6=q_{1a}^TC\gamma_{\mu}q_{2b}(\bar{q}_{3a}\gamma^{\mu}\gamma_5C\bar{q}^T_{4b}+\bar{q}_{3b}\gamma^{\mu}
\gamma_5C\bar{q}^T_{4a})-q_{1a}^TC\gamma_{\mu}\gamma_5q_{2b}(\bar{q}_{3a}\gamma^{\mu}C\bar{q}^T_{4b}+\bar{q}_{3b}\gamma^{\mu}C\bar{q}^T_{4a}) \, , \\
\eta^{(M)}_{2} &=&
A_3-P_3=q_{1a}^TC\gamma_{\mu}q_{2b}(\bar{q}_{3a}\gamma^{\mu}\gamma_5C\bar{q}^T_{4b}-\bar{q}_{3b}\gamma^{\mu}
\gamma_5C\bar{q}^T_{4a})-q_{1a}^TC\gamma_{\mu}\gamma_5q_{2b}(\bar{q}_{3a}\gamma^{\mu}C\bar{q}^T_{4b}-\bar{q}_{3b}\gamma^{\mu}C\bar{q}^T_{4a}) \, , \label{equ:current}\\
\nonumber \eta^{(A)} &=&
S_3-V_3=q_{1a}^TCq_{2b}(\bar{q}_{3a}\gamma_5C\bar{q}^T_{4b}-\bar{q}_{3b}\gamma_5C\bar{q}^T_{4a})-q_{1a}^TC\gamma_5q_{2b}(\bar{q}_{3a}C\bar{q}^T_{4b}-\bar{q}_{3b}C\bar{q}^T_{4a})
\, .
\end{eqnarray}
Adding different quauk contents as shown in Eq.~(\ref{content}),
there are altogether seven independent currents as shown :
\begin{enumerate}
\item For $6_F \otimes \bar{6}_F~(S)$:
\begin{eqnarray}
\nonumber \eta_1&=&S_6(qq\bar q \bar q)-V_6(qq\bar q \bar q) =
u^T_aCd_b(\bar{u}_a\gamma_5C\bar{d}^T_b+\bar{u}_b\gamma_5C\bar{d}^T_a)-
u^T_aC\gamma_5d_b(\bar{u}_aC\bar{d}^T_b+\bar{u}_bC\bar{d}^T_a) \, ,
\\
\eta_2&=&S_6(qs\bar q \bar s)-V_6(qs\bar q \bar s) =
u^T_aCs_b(\bar{u}_a\gamma_5C\bar{s}^T_b+\bar{u}_b\gamma_5C\bar{s}^T_a)-
u^T_aC\gamma_5s_b(\bar{u}_aC\bar{s}^T_b+\bar{u}_bC\bar{s}^T_a) \, ,
\end{eqnarray}
where $\eta_1$ belongs to the $\mathbf {27_F}$ representation and
contains up and down quarks only while $\eta_2$ belongs to the
$\mathbf {8_F}$ representation and contains one $s\bar s$ quark
pair.
\\

\item For $(\bar{3}_F \otimes \bar{6}_F) \oplus (6_F \otimes
3_F)~(M)$:
\begin{eqnarray}
\nonumber \eta_3&=&A_6(qq\bar q \bar q)-P_6(qq\bar q \bar q) =
u^T_aC\gamma_{\mu}d_b(\bar{u}_a\gamma^{\mu}\gamma_5C\bar{d}^T_b+\bar{u}_b\gamma^{\mu}
\gamma_5C\bar{d}^T_a)-u^T_aC\gamma_{\mu}\gamma_5d_b(\bar{u}_a\gamma^{\mu}C\bar{d}^T_b+\bar{u}_b
\gamma^{\mu}C\bar{d}^T_a) \, ,
\\ \nonumber
\eta_4&=&A_6(qs\bar q\bar s)-P_6(qs\bar q\bar s) =
u^T_aC\gamma_{\mu}s_b(\bar{u}_a\gamma^{\mu}\gamma_5C\bar{s}^T_b+\bar{u}_b\gamma^{\mu}
\gamma_5C\bar{s}^T_a)-u^T_aC\gamma_{\mu}\gamma_5s_b(\bar{u}_a\gamma^{\mu}C\bar{s}^T_b+\bar{u}_b
\gamma^{\mu}C\bar{s}^T_a) \, ,
\\
\eta_5&=&A_3(qq\bar q \bar q)-P_3(qq\bar q \bar q) =
u^T_aC\gamma_{\mu}d_b(\bar{u}_a\gamma^{\mu}C\bar{d}^T_b-\bar{u}_b
\gamma^{\mu}C\bar{d}^T_a)-u^T_aC\gamma_{\mu}\gamma_5d_b(\bar{u}_a\gamma^{\mu}C\bar{d}^T_b-\bar{u}_b\gamma^{\mu}
C\bar{d}^T_a) \, ,
\\
\nonumber \eta_6&=&A_3(qs\bar q\bar s)-P_3(qs\bar q\bar s) =
u^T_aC\gamma_{\mu}s_b(\bar{u}_a\gamma^{\mu}C\bar{s}^T_b-\bar{u}_b
\gamma^{\mu}C\bar{s}^T_a)-u^T_aC\gamma_{\mu}\gamma_5s_b(\bar{u}_a\gamma^{\mu}C\bar{s}^T_b-\bar{u}_b\gamma^{\mu}
C\bar{s}^T_a)\, ,
\end{eqnarray}
where $\eta_3$ and $\eta_5$ belong to the $\mathbf {\bar{10}_F}$
representation and contain only u, d quarks while $\eta_4$ and
$\eta_6$ belong to the $\mathbf {8_F}$ representation and contain
one $s\bar s$ quark pair.
\\

\item For $\bar{3}_F \otimes 3_F~(A)$:
\begin{eqnarray}
\eta_7=S_3(qs\bar q\bar s)-V_3(qs\bar q\bar s) =
u^T_aCs_b(\bar{u}_a\gamma_5C\bar{s}^T_b-\bar{u}_b\gamma_5C\bar{s}^T_a)-
u^T_aC\gamma_5s_b(\bar{u}_aC\bar{s}^T_b-\bar{u}_bC\bar{s}^T_a) \, .
\end{eqnarray}
where $\eta_7$ belongs to the  $\mathbf {8_F}$ and contains one
$s\bar s$ quark pair.
\end{enumerate}
It is understood that there exists the other part $\pm [u
\leftrightarrow d]$ in the expressions of $\eta_{2,4,6,7}$.

\subsection{The $J^{PC}=0^{+-}$ Tetraquark Currents}\label{sec:0+-}

Now we move on to the $J^{PC}=0^{+-}$ case. There are also ten
independent scalar tetraquark currents without derivative:
\begin{eqnarray}
\nonumber &&
S'_6=q_{1a}^TCq_{2b}(\bar{q}_{3a}C\bar{q}^T_{4b}+\bar{q}_{3b}C\bar{q}^T_{4a})
\, , \\ \nonumber &&
V'_6=q_{1a}^T\gamma_{\mu}Cq_{2b}(\bar{q}_{3a}C\gamma^{\mu}\bar{q}^T_{4b}+\bar{q}_{3b}C\gamma^{\mu}\bar{q}^T_{4a})
\, ,
\\ \nonumber &&
T'_6=q_{1a}^T\sigma_{\mu\nu}Cq_{2b}(\bar{q}_{3a}C\sigma^{\mu\nu}\bar{q}^T_{4b}+\bar{q}_{3b}C\sigma^{\mu\nu}
\bar{q}^T_{4a}) \, ,
\\ \nonumber &&
A'_6=q_{1a}^T\gamma_{\mu}\gamma_5Cq_{2b}(\bar{q}_{3a}C\gamma^{\mu}\gamma_5\bar{q}^T_{4b}+\bar{q}_{3b}C\gamma^{\mu}\gamma_5\bar{q}^T_{4a}) \, , \\
\nonumber &&
P'_6=q_{1a}^T\gamma_5Cq_{2b}(\bar{q}_{3a}C\gamma_5\bar{q}^T_{4b}+\bar{q}_{3b}C\gamma_5\bar{q}^T_{4a})
\, , \\&&
S'_3=q_{1a}^TCq_{2b}(\bar{q}_{3a}C\bar{q}^T_{4b}-\bar{q}_{3b}C\bar{q}^T_{4a})
\, , \\ \nonumber &&
V'_3=q_{1a}^T\gamma_{\mu}Cq_{2b}(\bar{q}_{3a}C\gamma^{\mu}\bar{q}^T_{4b}-\bar{q}_{3b}C\gamma^{\mu}\bar{q}^T_{4a})
\, ,
\\ \nonumber &&
T'_3=q_{1a}^T\sigma_{\mu\nu}Cq_{2b}(\bar{q}_{3a}C\sigma^{\mu\nu}\bar{q}^T_{4b}-\bar{q}_{3b}C\sigma^{\mu\nu}
\bar{q}^T_{4a}) \, ,
\\ \nonumber &&
A'_3=q_{1a}^T\gamma_{\mu}\gamma_5Cq_{2b}(\bar{q}_{3a}C\gamma^{\mu}\gamma_5\bar{q}^T_{4b}-\bar{q}_{3b}C\gamma^{\mu}\gamma_5\bar{q}^T_{4a}) \, , \\
\nonumber &&
P'_3=q_{1a}^T\gamma_5Cq_{2b}(\bar{q}_{3a}C\gamma_5\bar{q}^T_{4b}-\bar{q}_{3b}C\gamma_5\bar{q}^T_{4a})
\, . \label{currents1}
\end{eqnarray}
The flavor structure is again fixed due to the Pauli principle. To
have a charge-conjugation parity, we fix the quark contents to be:
$q_1 = q_3$ and $q_2 = q_4$ (or $q_1 = q_4$ and $q_2 = q_3$).
After performing the charge-conjugation transformation, we find
that they all have an even charge-conjugation parity, for example:
\begin{equation}
\mathbb{C} S'_6 \mathbb{C}^{-1}= + S'_6 \, .
\end{equation}
Therefore, the $J^{PC}=0^{+-}$ tetraquark interpolating currents
without derivatives do not exist.

%
\section{The Spectral Density}\label{sec:ope}

We consider the two-point correlation function in the framework of
QCD sum rule ~\cite{Shifman:1978bx,Reinders:1984sr}:
\begin{equation}
\Pi(q^{2})\equiv\int d^4
xe^{iqx}\langle0|T\eta(x)\eta^\dag(0)|0\rangle, \label{equ:po}
\end{equation}
where $\eta$ is an interpolating current. We can calculate
$\Pi(q^{2})$ at the quark gluon level using the propagator:
\begin{eqnarray}
\nonumber iS^{ab}_q &\equiv&
\langle0|T[q^a(x)\bar{q}^b(0)]|0\rangle
\\ \nonumber
&=&  \frac{i\delta^{ab}}{2\pi^2x^4}\hat{x}
+\frac{i}{32\pi^2}\frac{\lambda^n_{ab}}{2}gG_{\mu\nu}^n\frac{1}{x^2}(\sigma^{\mu\nu}\hat{x}+\hat{x}\sigma^{\mu\nu})
-\frac{\delta^{ab}}{12}\langle\bar{q}q\rangle
\\
&&+ \frac{\delta^{ab}x^2}{192}\langle g_s \bar{q}\sigma Gq\rangle
-\frac{m_q\delta^{ab}}{4\pi^2x^2}+\frac{i\delta^{ab}m_q\langle\bar{q}q\rangle}{48}\hat{x}
+\frac{i\delta^{ab}m_q^2\langle\bar{q}q\rangle}{8\pi^2x^2}\hat{x},
\end{eqnarray}
where $\hat{x}\equiv\gamma_{\mu}x^{\mu}$. With the dispersion
relation $\Pi(q^{2})$ is related to the observable at the hadron
level
\begin{equation}
\Pi(p^2)=\int_{0}^{\infty}\frac{\rho(s)}{s-p^2-i\varepsilon}ds,\label{equ:pq}
\end{equation}
where
\begin{equation}
\rho(s)\equiv\sum_{n}\delta(s-M^{2}_{n})\langle0|\eta|n\rangle\langle
n|\eta^{\dag}|0\rangle
=f^{2}_{X}\delta(s-M^{2}_{X})+\mbox{continuum} \; .
\label{equ:rho}
\end{equation}
Here, the usual pole plus continuum parametrization of the hadronic
spectral density is adopted. Up to dimension 12, the spectral
density $\rho_i(s)$ at the quark and gluon level reads:
\begin{eqnarray}
\nonumber
\rho_1(s)&=&\frac{s^4}{15360\pi^6}-\frac{m_q^2}{192\pi^6}s^3-(\frac{\langle
g_s^2GG\rangle}{3072\pi^6}-\frac{m_q
\langle\bar{q}q\rangle}{24\pi^4})s^2 +[\frac{\langle g_s^2GG\rangle
m_q^2}{256\pi^6}+\frac{\langle g_s^3 f GGG\rangle}{768\pi^6}(3\ln
(\frac{s}{\tilde{\mu}^2})-5)]s
\\
 && - (\frac{3 m_q^2
\langle\bar{q}q\rangle^2}{2\pi^2}+\frac{\langle g_s^2GG\rangle m_q
\langle\bar{q}q\rangle}{192\pi^4})+(\frac{16}{9}m_q
\langle\bar{q}q\rangle^3-\frac{1}{\pi^2}m_q^2 \langle\bar{q}q\rangle
\langle g_s \bar{q}\sigma Gq\rangle)\delta(s) \, ,
\end{eqnarray}
\begin{eqnarray}
\nonumber
\rho_2(s)&=&\frac{s^4}{15360\pi^6}-\frac{m_s^2}{384\pi^6}s^3+(\frac{m_s^4}{64\pi^6}+\frac{m_s
\langle\bar{s}s\rangle}{48\pi^4}-\frac{\langle
g_s^2GG\rangle}{3072\pi^6})s^2
\\ \nonumber
 &&+[\frac{\langle g_s^3 f
GGG\rangle}{768\pi^6}(3\ln (\frac{s}{\tilde{\mu}^2})-5)-(\frac{m_s^3
\langle\bar{s}s\rangle}{8\pi^4}-\frac{m_s^2 \langle
g_s^2GG\rangle}{512\pi^6})]s
\\
&& +(\frac{m_s^2 \langle\bar{s}s\rangle^2}{12\pi^2}
 - \frac{m_s^2 \langle\bar{q}q\rangle^2}{3\pi^2}-\frac{m_s
\langle\bar{s}s\rangle \langle
g_s^2GG\rangle}{384\pi^4})-(\frac{m_s^2 \langle\bar{u}u\rangle
\langle g_s \bar{q}\sigma Gq\rangle}{6\pi^2}-\frac{8}{9}m_s
\langle\bar{s}s\rangle \langle\bar{q}q\rangle^2)\delta(s) \, ,
\end{eqnarray}
\begin{eqnarray}
\nonumber
\rho_3(s)&=&\frac{s^4}{3840\pi^6}-\frac{m_q^2}{48\pi^6}s^3+(\frac{5\langle
g_s^2GG\rangle}{1536\pi^6}+\frac{m_q
\langle\bar{q}q\rangle}{6\pi^4})s^2+[\frac{\langle g_s^3 f
GGG\rangle}{192\pi^6}(3\ln
(\frac{s}{\tilde{\mu}^2})-5)-\frac{5\langle g_s^2GG\rangle
m_q^2}{128\pi^6}]s
\\
 && - (\frac{6 m_q^2
\langle\bar{q}q\rangle^2}{\pi^2}-\frac{5 \langle g_s^2GG\rangle m_q
\langle\bar{q}q\rangle}{96\pi^4}) +(\frac{64}{9}m_q
\langle\bar{q}q\rangle^3-\frac{4}{\pi^2}m_q^2 \langle\bar{q}q\rangle
\langle g_s \bar{q}\sigma Gq\rangle)\delta(s) \, ,
\end{eqnarray}
\begin{eqnarray}
\nonumber
\rho_4(s)&=&\frac{s^4}{3840\pi^6}-\frac{m_s^2}{96\pi^6}s^3+(\frac{m_s^4}{16\pi^6}+\frac{m_s
\langle\bar{s}s\rangle}{12\pi^4}+\frac{5\langle
g_s^2GG\rangle}{1536\pi^6})s^2
\\ \nonumber
 &&+[\frac{\langle g_s^3 f
GGG\rangle}{192\pi^6}(3\ln (\frac{s}{\tilde{\mu}^2})-5)-(\frac{m_s^3
\langle\bar{s}s\rangle}{2\pi^4}+\frac{5m_s^2 \langle
g_s^2GG\rangle}{256\pi^6})]s
\\
&& +(\frac{m_s^2 \langle\bar{s}s\rangle^2}{3\pi^2}
 -\frac{4m_s^2 \langle\bar{q}q\rangle^2}{3\pi^2}+ \frac{5m_s
\langle\bar{s}s\rangle \langle
g_s^2GG\rangle}{192\pi^4})-(\frac{2m_s^2 \langle\bar{u}u\rangle
\langle g_s \bar{q}\sigma Gq\rangle}{3\pi^2}-\frac{32}{9}m_s
\langle\bar{s}s\rangle \langle\bar{q}q\rangle^2)\delta(s) \, ,
\end{eqnarray}
\begin{eqnarray}
\nonumber
\rho_5(s)&=&\frac{s^4}{7680\pi^6}-\frac{m_q^2}{96\pi^6}s^3+(\frac{\langle
g_s^2GG\rangle}{1536\pi^6}+\frac{m_q
\langle\bar{q}q\rangle}{12\pi^4})s^2+[\frac{\langle g_s^3 f
GGG\rangle}{384\pi^6}(3\ln
(\frac{s}{\tilde{\mu}^2})-5)-\frac{\langle g_s^2GG\rangle
m_q^2}{128\pi^6}]s
\\
&& - (\frac{3 m_q^2 \langle\bar{q}q\rangle^2}{\pi^2}-\frac{ \langle
g_s^2GG\rangle m_q \langle\bar{q}q\rangle}{96\pi^4})
+(\frac{32}{9}m_q \langle\bar{q}q\rangle^3-\frac{2}{\pi^2}m_q^2
\langle\bar{q}q\rangle \langle g_s \bar{q}\sigma Gq\rangle)\delta(s)
\, ,
\end{eqnarray}
\begin{eqnarray}
\nonumber
\rho_6(s)&=&\frac{s^4}{7680\pi^6}-\frac{m_s^2}{192\pi^6}s^3+(\frac{m_s^4}{32\pi^6}+\frac{m_s
\langle\bar{s}s\rangle}{24\pi^4}+\frac{\langle
g_s^2GG\rangle}{1536\pi^6})s^2
\\ \nonumber
 &&+[\frac{\langle g_s^3 f
GGG\rangle}{384\pi^6}(3\ln (\frac{s}{\tilde{\mu}^2})-5)-(\frac{m_s^3
\langle\bar{s}s\rangle}{4\pi^4}+\frac{m_s^2 \langle
g_s^2GG\rangle}{256\pi^6})]s
\\
&&+(\frac{m_s^2 \langle\bar{s}s\rangle^2}{6\pi^2} - \frac{2m_s^2
\langle\bar{q}q\rangle^2}{3\pi^2}+\frac{m_s \langle\bar{s}s\rangle
\langle g_s^2GG\rangle}{192\pi^4}) -(\frac{m_s^2
\langle\bar{u}u\rangle \langle g_s \bar{q}\sigma
Gq\rangle}{3\pi^2}-\frac{16}{9}m_s \langle\bar{s}s\rangle
\langle\bar{q}q\rangle^2)\delta(s) \, , \label{pi6}
\end{eqnarray}
\begin{eqnarray}
\nonumber
\rho_7(s)&=&\frac{s^4}{30720\pi^6}-\frac{m_s^2}{768\pi^6}s^3+(\frac{m_s^4}{128\pi^6}+\frac{m_s
\langle\bar{s}s\rangle}{96\pi^4}+\frac{\langle
g_s^2GG\rangle}{3072\pi^6})s^2
\\ \nonumber
&&+[\frac{\langle g_s^3 f GGG\rangle}{1536\pi^6}(3\ln
(\frac{s}{\tilde{\mu}^2})-5) -(\frac{m_s^3
\langle\bar{s}s\rangle}{16\pi^4}+\frac{m_s^2 \langle
g_s^2GG\rangle}{512\pi^6})]s
\\
&&+(\frac{m_s^2
\langle\bar{s}s\rangle^2}{24\pi^2}- \frac{m_s^2
\langle\bar{q}q\rangle^2}{6\pi^2}+ \frac{m_s \langle\bar{s}s\rangle
\langle g_s^2GG\rangle}{384\pi^4})-(\frac{m_s^2
\langle\bar{u}u\rangle \langle g_s \bar{q}\sigma
Gq\rangle}{12\pi^2}-\frac{4}{9}m_s \langle\bar{s}s\rangle
\langle\bar{q}q\rangle^2)\delta(s) \, .
\end{eqnarray}
It is interesting to note several important features of the above
spectral densities:
\begin{itemize}
\item First the special Lorentz structure of the $J^{PC}=0^{--}$
interpolating currents forbids the appearance of the four-quark
type of condensates $\langle\bar{q}q\rangle^2$,
$\langle\bar{q}q\rangle$ $\langle g_s \bar{q}\sigma Gq\rangle$ and
$\langle g_s \bar{q}\sigma Gq\rangle^2$. Usually these terms play
an important role in the multiquark sum rules. The Feynman
diagrams for the dimension 10 condensate $\langle g_s
\bar{q}\sigma Gq\rangle^2$ are shown in Fig.
\ref{fig:mixed-condensate}.
%
\begin{figure}[hbtp]
\begin{center}
\begin{tabular}{llr}
\scalebox{0.75}{\includegraphics{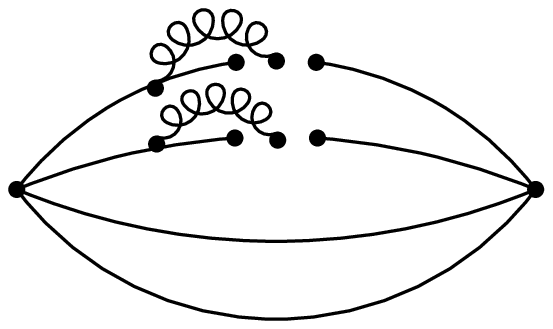}}&
\scalebox{0.75}{\includegraphics{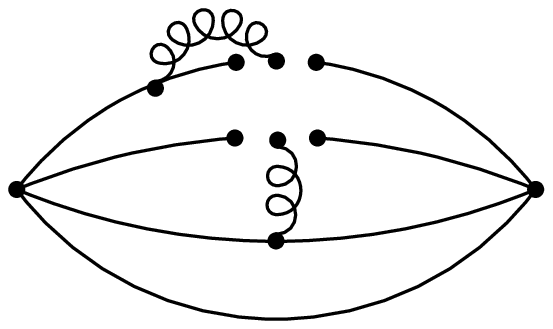}}&
\scalebox{0.75}{\includegraphics{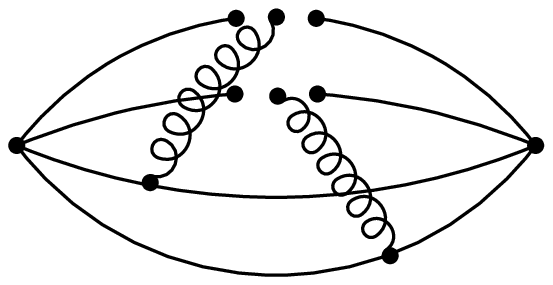}}
\end{tabular}
\caption{Feynman diagrams for the quark gluon mixed condensate.}
\label{fig:mixed-condensate}
\end{center}
\end{figure}
%

\item The dominant non-perturbative correction arises from the
gluon condensate, which is destructive for $\rho_{1-2}(s)$ and
constructive for $\rho_{3-7}(s)$. Moreover there are corrections
from the tri-gluon condensate $\langle g_s^3 f^{abc}
G^aG^bG^c\rangle$ as shown in Fig.\ref{fig:tri-gluon}. In the above
expressions we use the short-hand notation $\langle g_s^3 f
GGG\rangle$ to denote the tri-gluon condensate. There are three
types of Feynman diagrams. The first class of Feynman diagrams
vanishes because of the product of the color matrices. The second
class is proportional to $m_q$ and could be omitted in the chiral
limit. Only the third class leads to non-vanishing tri-gluon
correction. In fact the gluon condensates become the only power
corrections in the chiral limit.

\begin{figure}[hbtp]
\begin{center}
\begin{tabular}{llr}
\scalebox{0.75}{\includegraphics{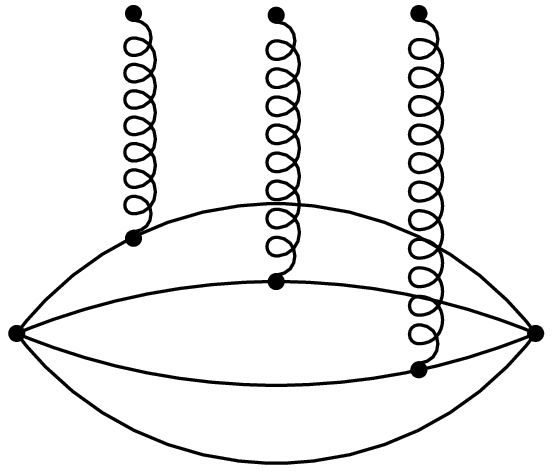}}&
\scalebox{0.75}{\includegraphics{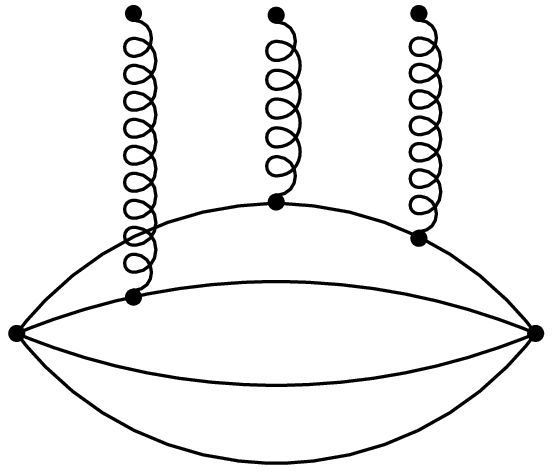}}&
\scalebox{0.75}{\includegraphics{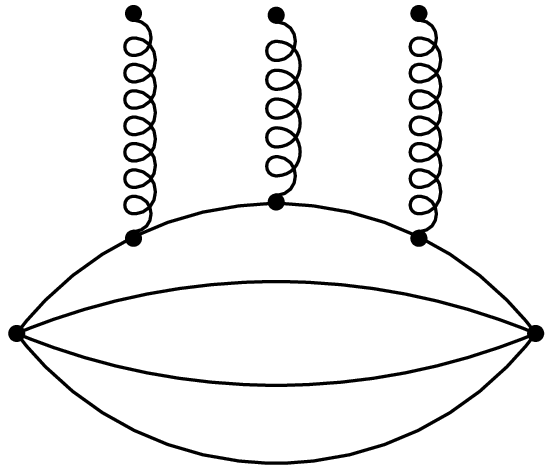}}
\end{tabular}
\caption{Feynman diagrams for the tri-gluon condensate.}
\label{fig:tri-gluon}
\end{center}
\end{figure}
%

\item The second term in each $\rho_i(s)$ is destructive, which
renders the spectral density negative when $s$ is small. This
$-m_q^2 s^3$ piece is an artefact of the expansion of the quark
propagator ${i\over {\hat p}-m_q}$ in terms of the quark mass $m_q$
perturbatively. Without making such an expansion, the perturbative
contribution to the spectral density is always positive-definite.
Such a destructive term will sometimes produce an artificial plateau
and stability window in the sum rule analysis, which must be
removed.

\item Although the tree-level four-quark condensate vanishes, one
may wonder whether the four-quark condensate $ g_s^2\langle
\bar{q}q\rangle^2$  plays a role since the latter is very important
in the $q\bar q$ meson sum rules
\cite{Shifman:1978bx,Reinders:1984sr}. Two types of Feynman diagrams
could produce such a correction. The first class of Feynman diagrams
is very similar to that in the $q\bar q$ meson case where a gluon
propagator is attached between two-quark condensates, as
Fig.\ref{fig:4-conden} shown. It's easy to check that they vanish
due to the special Lorentz structure of the correlation function.
%
%
\begin{figure}[hbtp]
\begin{center}
\begin{tabular}{lr}
\scalebox{0.75}{\includegraphics{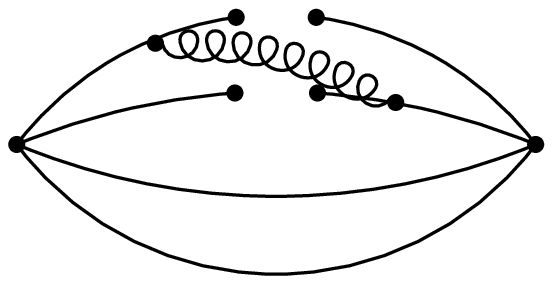}}&
\scalebox{0.75}{\includegraphics{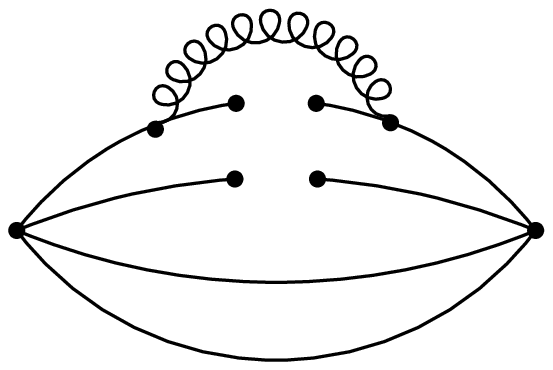}}
\end{tabular}
\caption{One set of Feynman diagrams for the four-quark
condensate.} \label{fig:4-conden}
\end{center}
\end{figure}
%
%
One of the second class of diagrams is shown in
Fig.\ref{fig:4-condensate}. In this case, we use the mesonic type
interpolating currents in the appendix\ref{app} to simplify the
derivation. After making Wick-contraction to the correlation
function,
\begin{eqnarray}\nonumber
\bar{\psi}_3(x)\Gamma'_1\psi_4(x)\bar{\psi}_1(x)\Gamma_1\psi_2(x)\bar{\psi}_1(z_1)
gt^a\gamma^{\mu}\psi_1(z_1)A_{\mu}^a(z_1)\bar{\psi}_2(z_2)gt^b\gamma^{\nu}\psi_2(z_2)
A_{\nu}^b(z_2)\bar{\psi}_2(y)\Gamma_2\psi_1(y)\bar{\psi}_4(y)\Gamma'_2\psi_3(y)
\end{eqnarray}
we get
\begin{eqnarray}\nonumber
Tr[-\Gamma'_1S_Q(x-y)\Gamma'_2S_Q(y-x)]
Tr[-S_Q(x-z_2)\gamma^{\nu}S_Q(z_2-y)\Gamma_2S_Q(y-z_1)\gamma^{\mu}S_Q(z_1-x)\Gamma_1\times
g_{\mu\nu}\times S_G(z_2-z_1)].
\end{eqnarray}
where $S_Q$ is the quark propagator and $S_G$ is the gluon
propagator. \{$\Gamma_1$,$\Gamma_2$\} could be either \{$I$,
$\gamma_5$\} or \{$\gamma_\alpha$, $\gamma_5\gamma_\alpha$\}.
$S_Q(y-z_1)\propto\langle\bar qq\rangle$. In fact, there would be
three $\gamma$-matrices or three $\gamma$-matrices plus $\gamma_5$
left in the latter trace. Therefore this piece also vanishes.
\begin{figure}[hbtp]
\begin{center}
\begin{tabular}{lr}
\scalebox{0.75}{\includegraphics{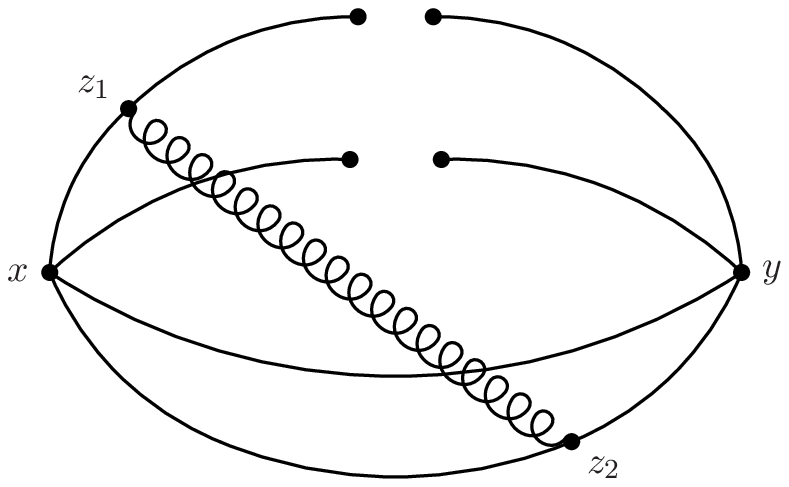}}
\end{tabular}
\caption{Feynman diagrams for the four-quark condensate.}
\label{fig:4-condensate}
\end{center}
\end{figure}
%

\end{itemize}

\section{Numerical Analysis}\label{sec:num}
%
In the chiral limit ($m_s=m_q=0$) the spectral density reads
\begin{eqnarray}
\nonumber \rho_{1-2}(s)&=&\frac{s^4}{15360\pi^6}-\frac{\langle
g_s^2GG\rangle}{3072\pi^6}s^2 +\frac{\langle g_s^3 f
GGG\rangle}{768\pi^6}(3\ln (\frac{s}{\tilde{\mu}^2})-5)s,
\\ \nonumber
\rho_{3-4}(s)&=&\frac{s^4}{3840\pi^6}-\frac{5\langle
g_s^2GG\rangle}{1536\pi^6}s^2 +\frac{\langle g_s^3 f
GGG\rangle}{192\pi^6}(3\ln (\frac{s}{\tilde{\mu}^2})-5)s,
\\ \nonumber
\rho_{5-6}(s)&=&\frac{s^4}{7680\pi^6}-\frac{\langle
g_s^2GG\rangle}{1536\pi^6}s^2 +\frac{\langle g_s^3 f
GGG\rangle}{384\pi^6}(3\ln (\frac{s}{\tilde{\mu}^2})-5)s,
\\
\rho_{7}(s)&=&\frac{s^4}{30720\pi^6}-\frac{\langle
g_s^2GG\rangle}{3072\pi^6}s^2 +\frac{\langle g_s^3 f
GGG\rangle}{1536\pi^6}(3\ln (\frac{s}{\tilde{\mu}^2})-5)s
\end{eqnarray}
where $\tilde{\mu}=1$ GeV. Requiring the pole contribution is
larger than $40\%$, one gets the upper bound $M^2_{\mbox{max}}$ of
the Borel parameter $M_B^2$. The convergence of the operator
expansion product leads to the lower bound $M^2_{\mbox{min}}$ of
the Borel parameter. In the present case, we require that the two
gluon condensate correction be less than one third of the
perturbative term and the tri-gluon condensate correction less
than one third of the gluon condensate correction. The working
region of $M_B^2$ in the sum rule analysis is [$M^2_{\mbox{min}}$,
$M^2_{\mbox{max}}$], which is dependent on the threshold $s_0$.

\begin{figure}[hbtp]
\begin{center}
\begin{tabular}{lr}
\scalebox{0.6}{\includegraphics{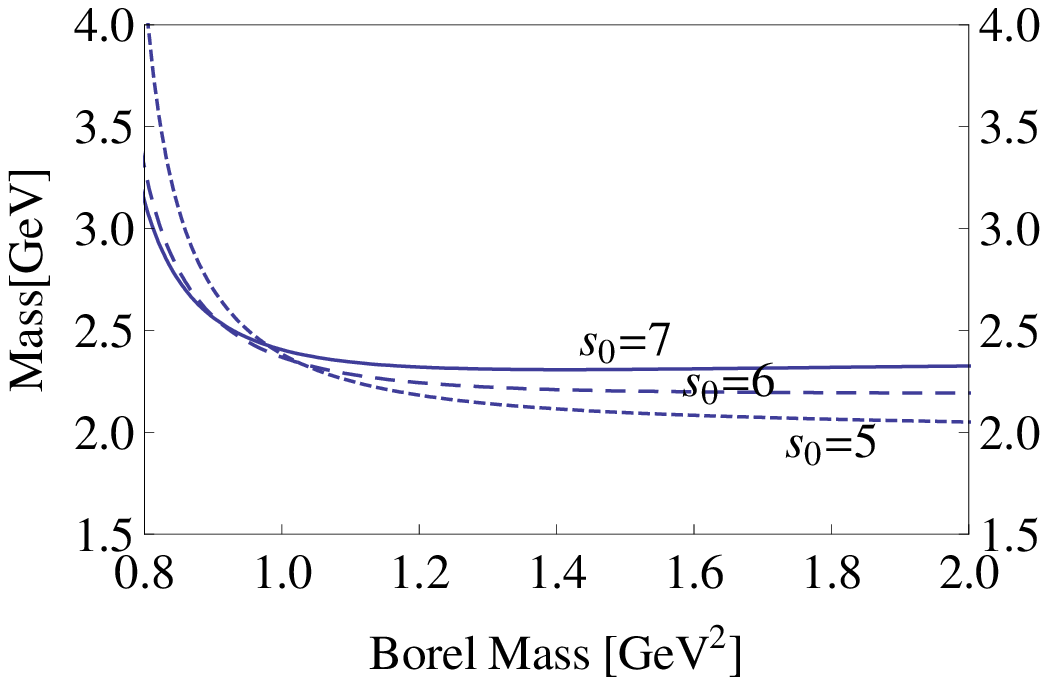}}&
\scalebox{0.6}{\includegraphics{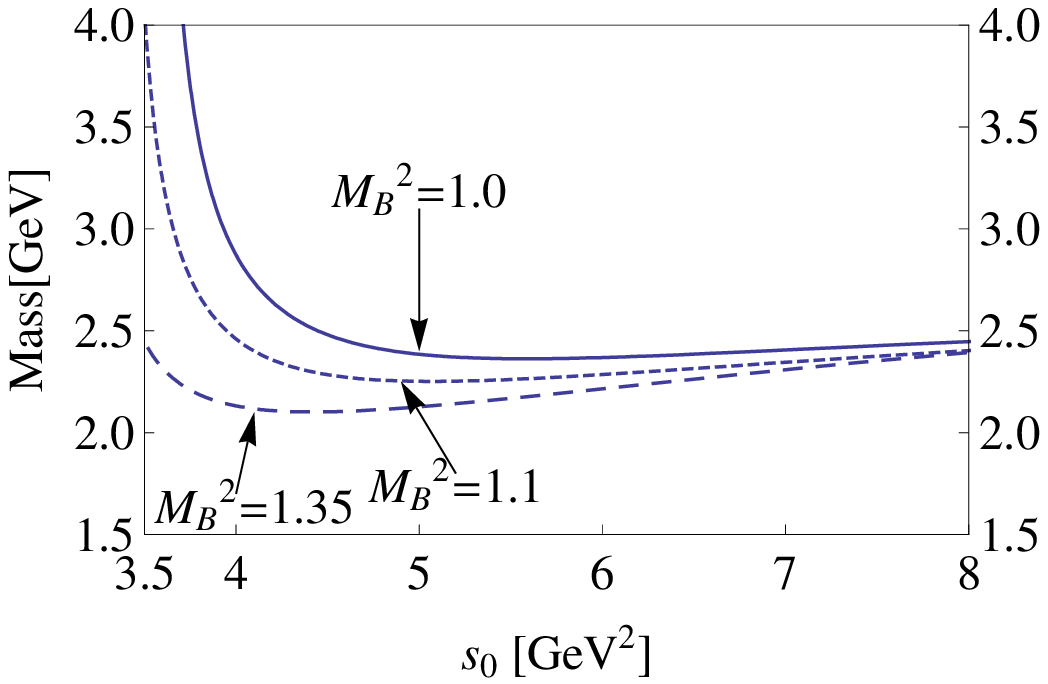}}
\end{tabular}
\caption{The variation of $M_X$ with $M^2_B$ (Left) and $s_0$
(Right) for the current $\eta_{1-2}$ using Ioffe's gluon
condensate values.} \label{fig:eta12Ioffe}
\end{center}
\end{figure}

\begin{figure}[hbtp]
\begin{center}
\begin{tabular}{lr}
\scalebox{0.6}{\includegraphics{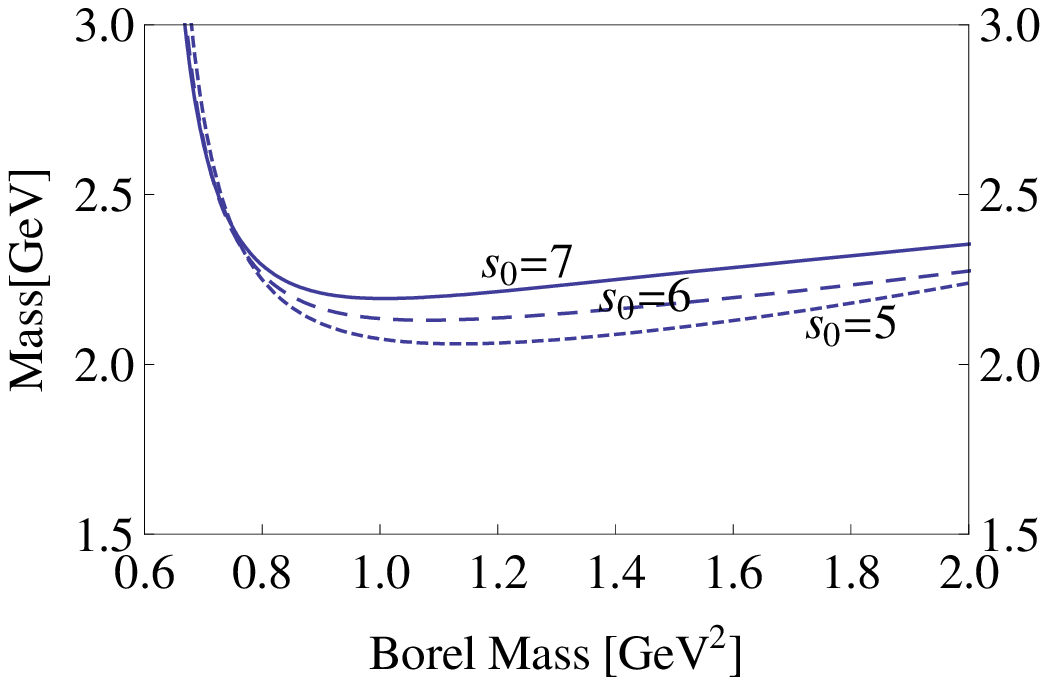}}&
\scalebox{0.6}{\includegraphics{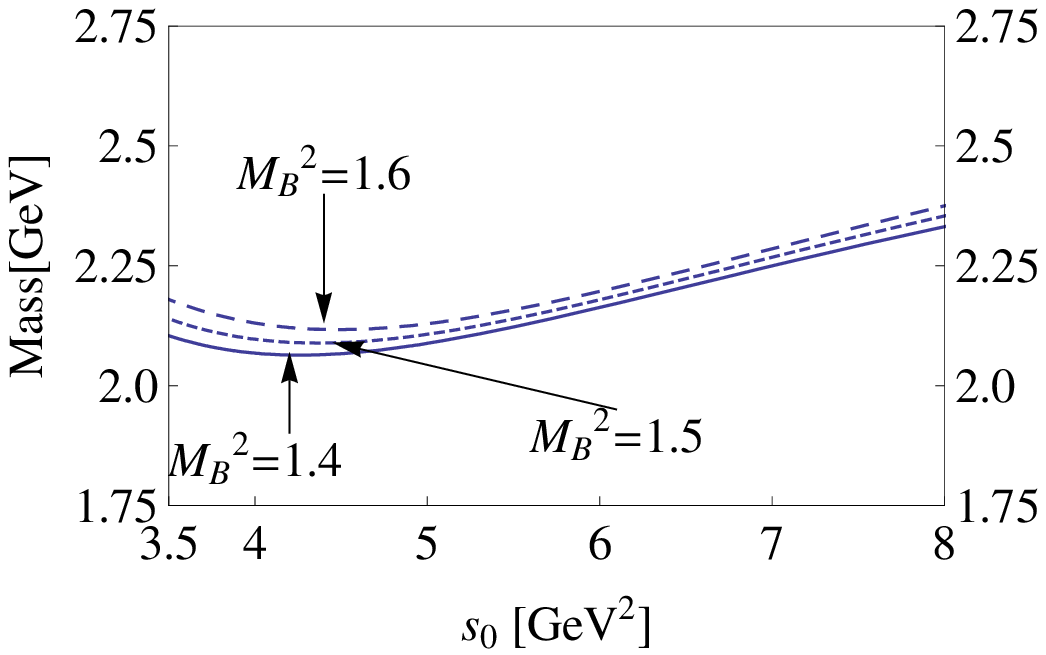}}
\end{tabular}
\caption{The variation of $M_X$ with $M^2_B$ (Left) and $s_0$
(Right) for the current $\eta_{3-4}$ using Ioffe's gluon
condensate values.} \label{fig:eta34Ioffe}
\end{center}
\end{figure}

\begin{figure}[hbtp]
\begin{center}
\begin{tabular}{lr}
\scalebox{0.6}{\includegraphics{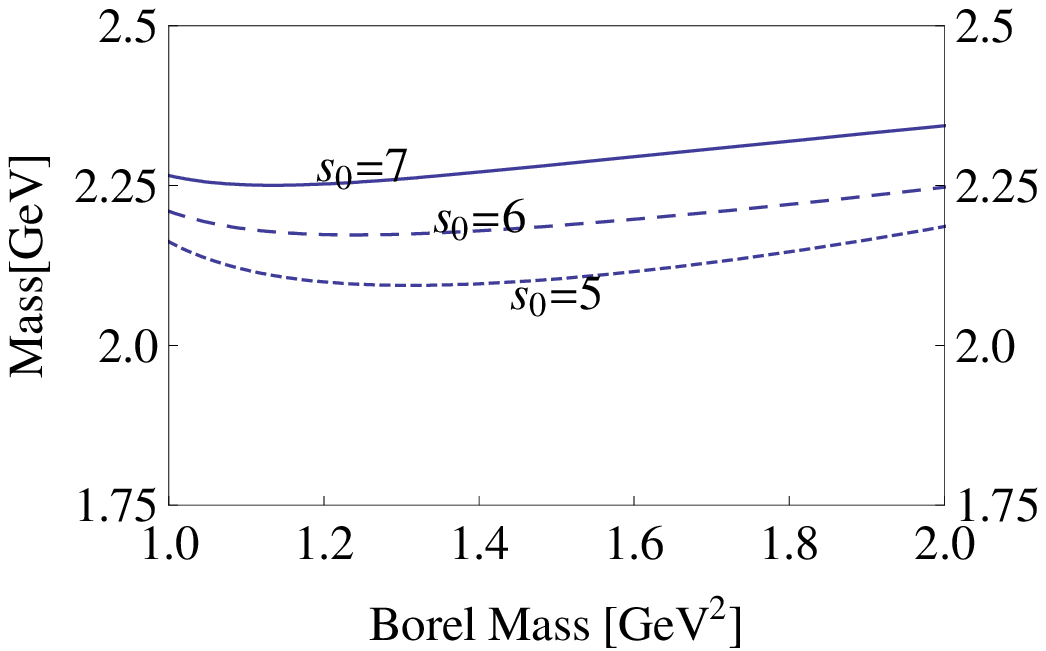}}&
\scalebox{0.6}{\includegraphics{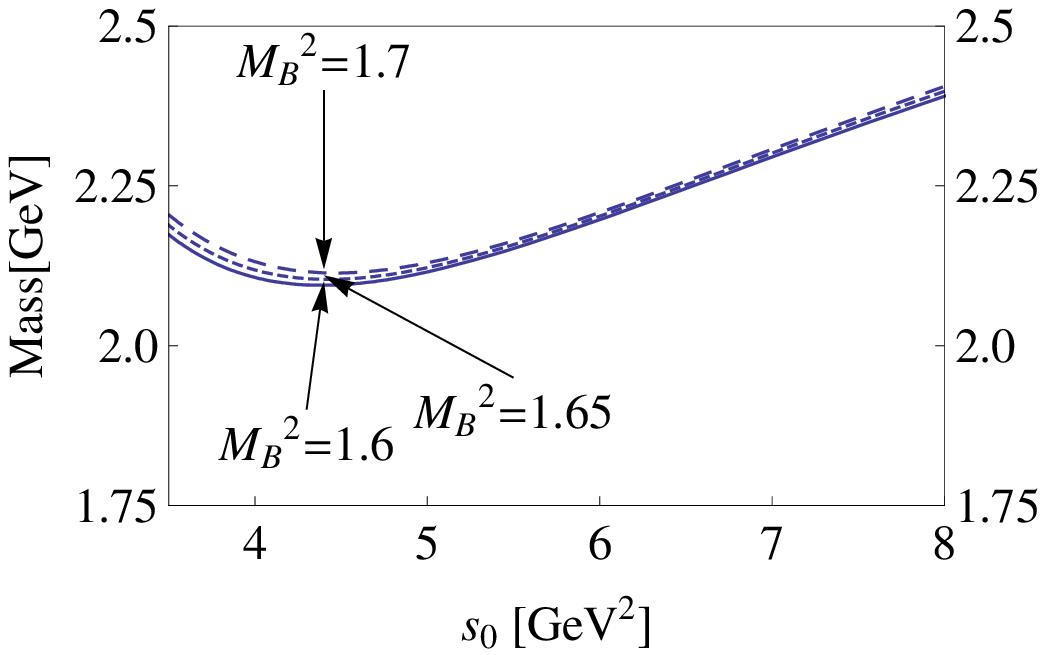}}
\end{tabular}
\caption{The variation of $M_X$ with $M^2_B$ (Left) and $s_0$
(Right) for the current $\eta_{5-6}$ using Ioffe's gluon
condensate values.} \label{fig:eta56Ioffe}
\end{center}
\end{figure}

\begin{figure}[hbtp]
\begin{center}
\begin{tabular}{lr}
\scalebox{0.6}{\includegraphics{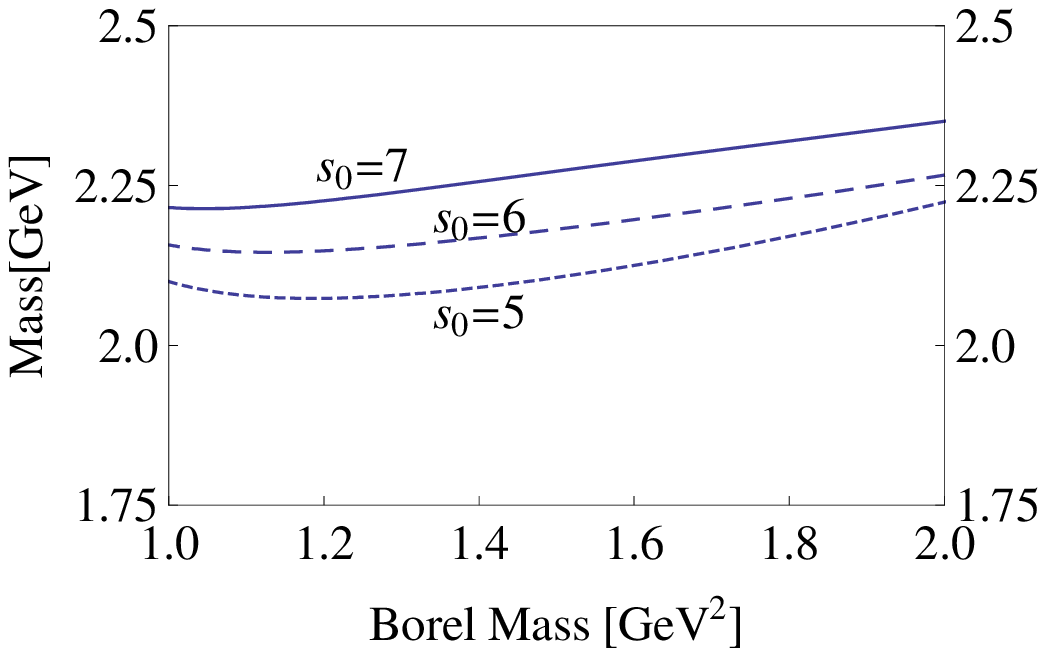}}&
\scalebox{0.6}{\includegraphics{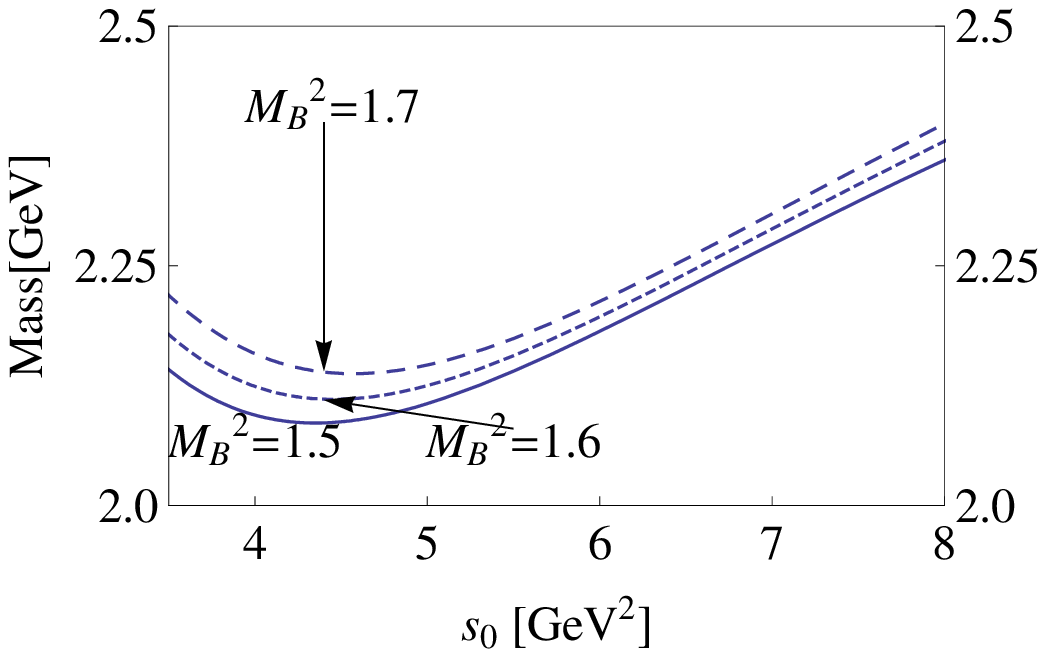}}
\end{tabular}
\caption{The variation of $M_X$ with $M^2_B$ (Left) and $s_0$
(Right) for the current $\eta_{7}$ using Ioffe's gluon condensate
values.} \label{fig:eta7Ioffe}
\end{center}
\end{figure}

\begin{figure}[hbtp]
\begin{center}
\begin{tabular}{lr}
\scalebox{0.6}{\includegraphics{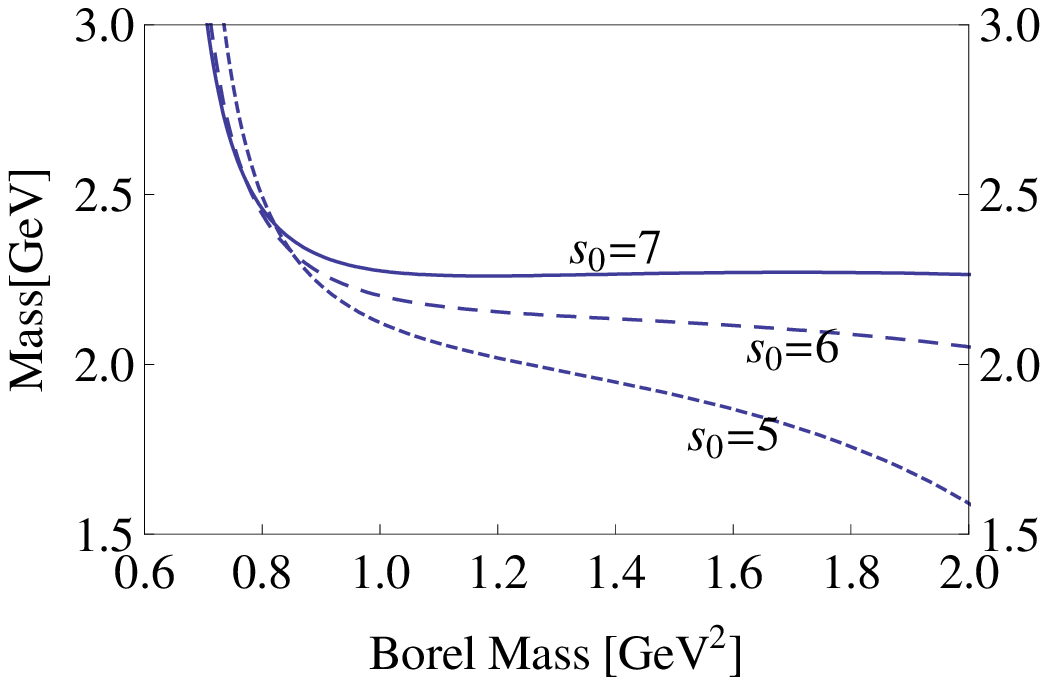}}&
\scalebox{0.6}{\includegraphics{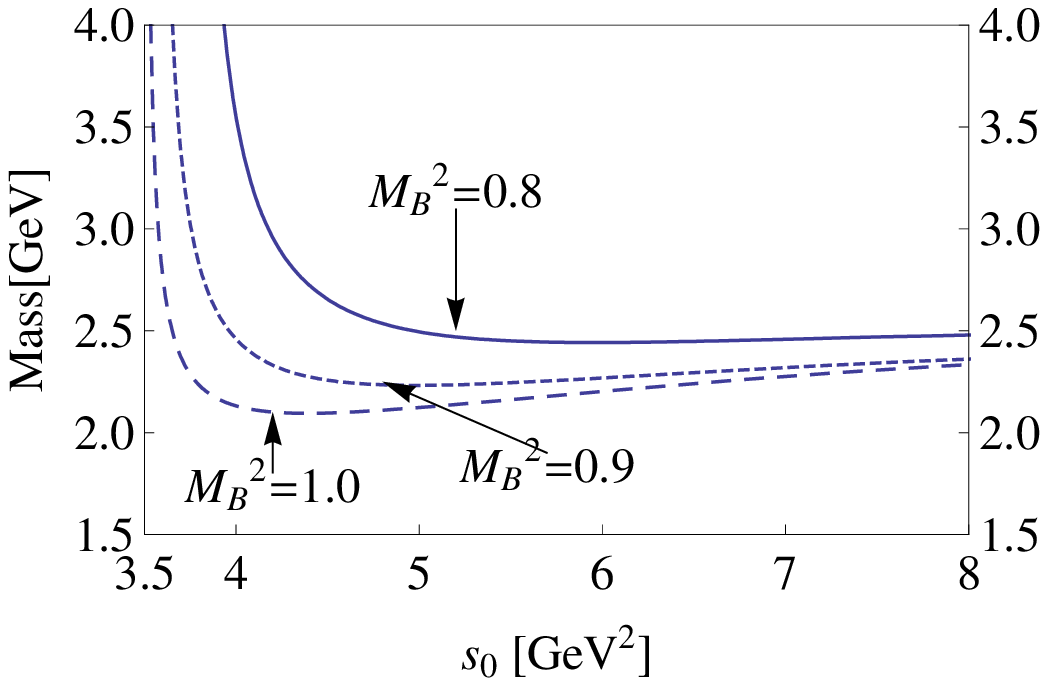}}
\end{tabular}
\caption{The variation of $M_X$ with $M^2_B$ (Left) and $s_0$
(Right) for the current $\eta_{1-2}$ using SVZ's gluon condensate
values.} \label{fig:eta12Shifman}
\end{center}
\end{figure}

\begin{figure}[hbtp]
\begin{center}
\begin{tabular}{lr}
\scalebox{0.6}{\includegraphics{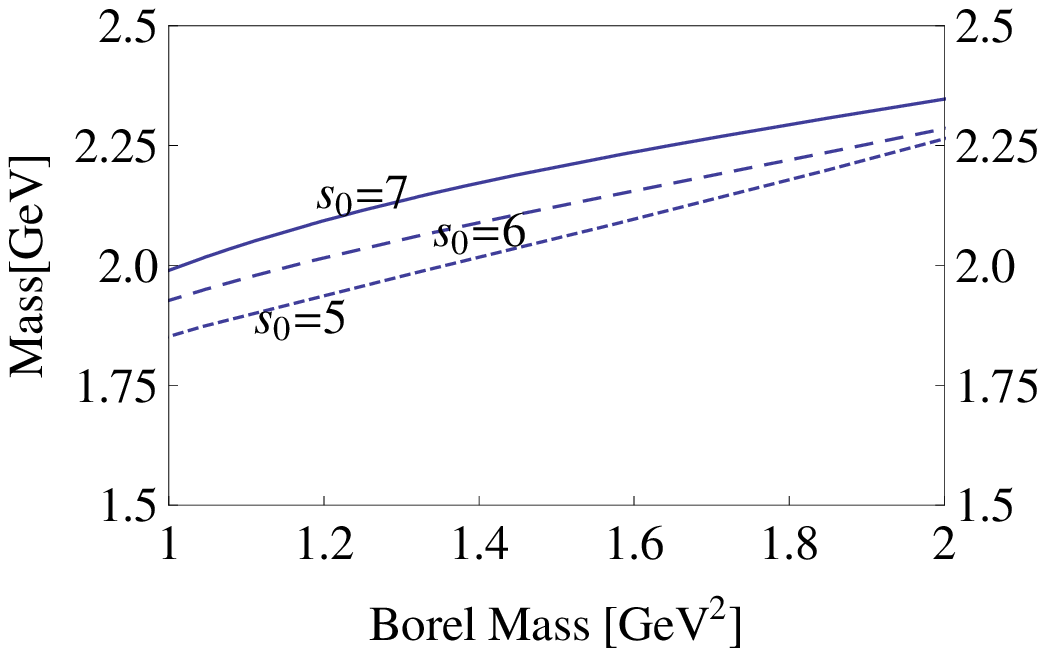}}&
\scalebox{0.6}{\includegraphics{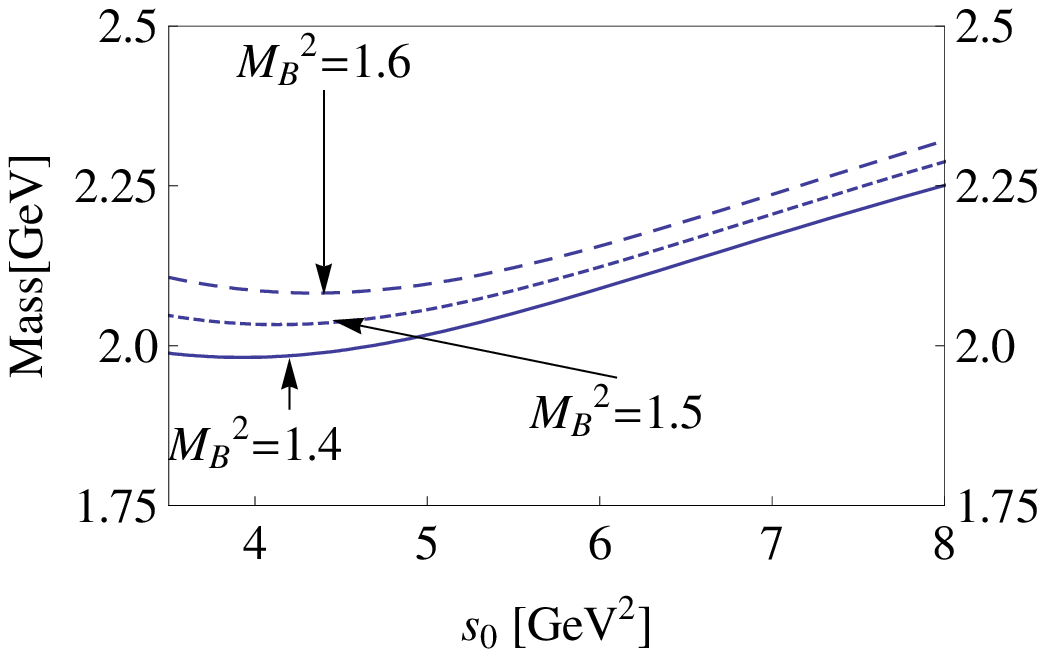}}
\end{tabular}
\caption{The variation of $M_X$ with $M^2_B$ (Left) and $s_0$
(Right) for the current $\eta_{3-4}$ using SVZ's gluon condensate
values.} \label{fig:eta34Shifman}
\end{center}
\end{figure}

\begin{figure}[hbtp]
\begin{center}
\begin{tabular}{lr}
\scalebox{0.6}{\includegraphics{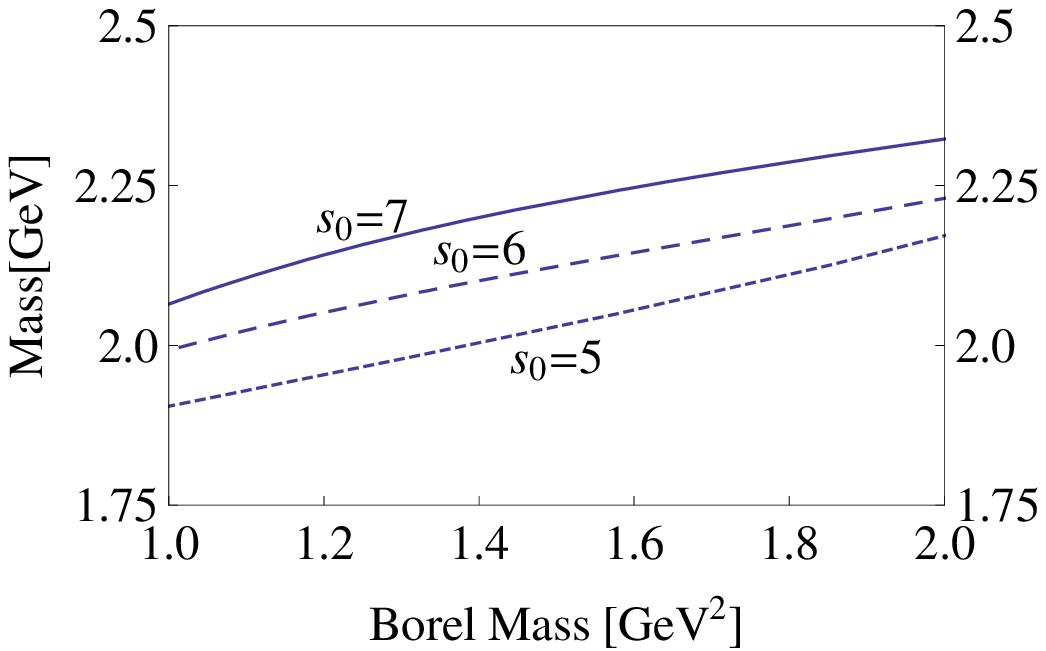}}&
\scalebox{0.6}{\includegraphics{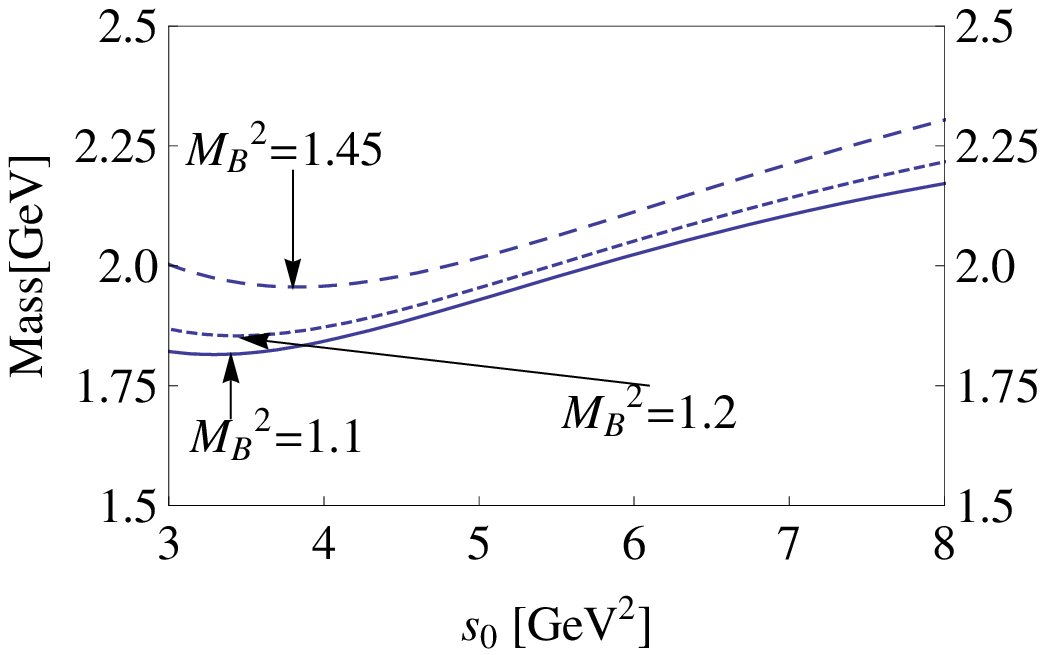}}
\end{tabular}
\caption{The variation of $M_X$ with $M^2_B$ (Left) and $s_0$
(Right) for the current $\eta_{5-6}$ using SVZ's gluon condensate
values.} \label{fig:eta56Shifman}
\end{center}
\end{figure}

\begin{figure}[hbtp]
\begin{center}
\begin{tabular}{lr}
\scalebox{0.6}{\includegraphics{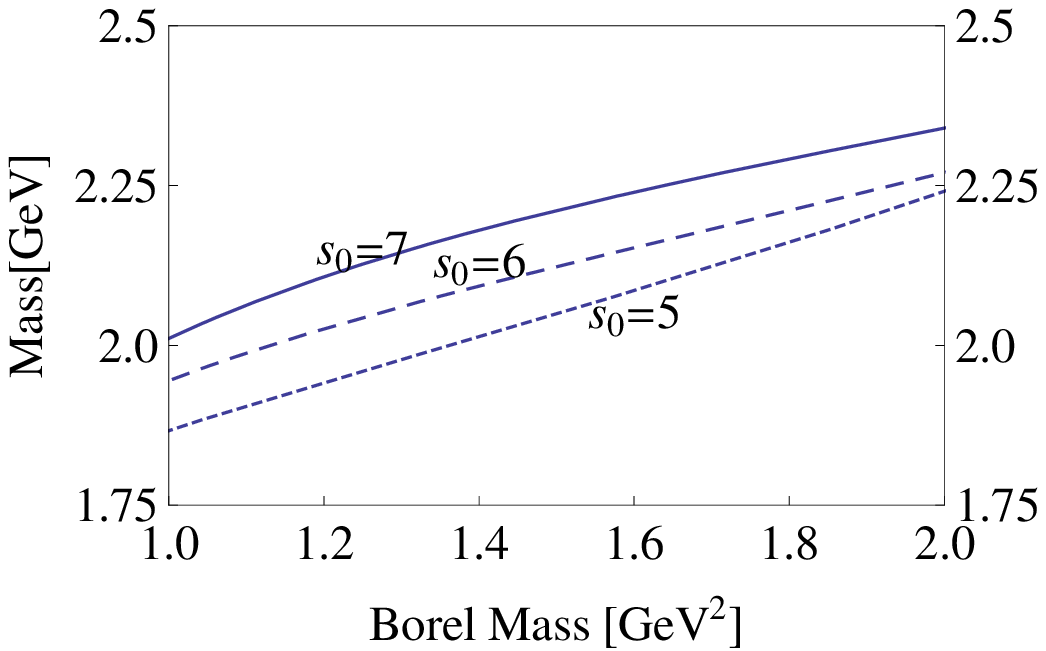}}&
\scalebox{0.6}{\includegraphics{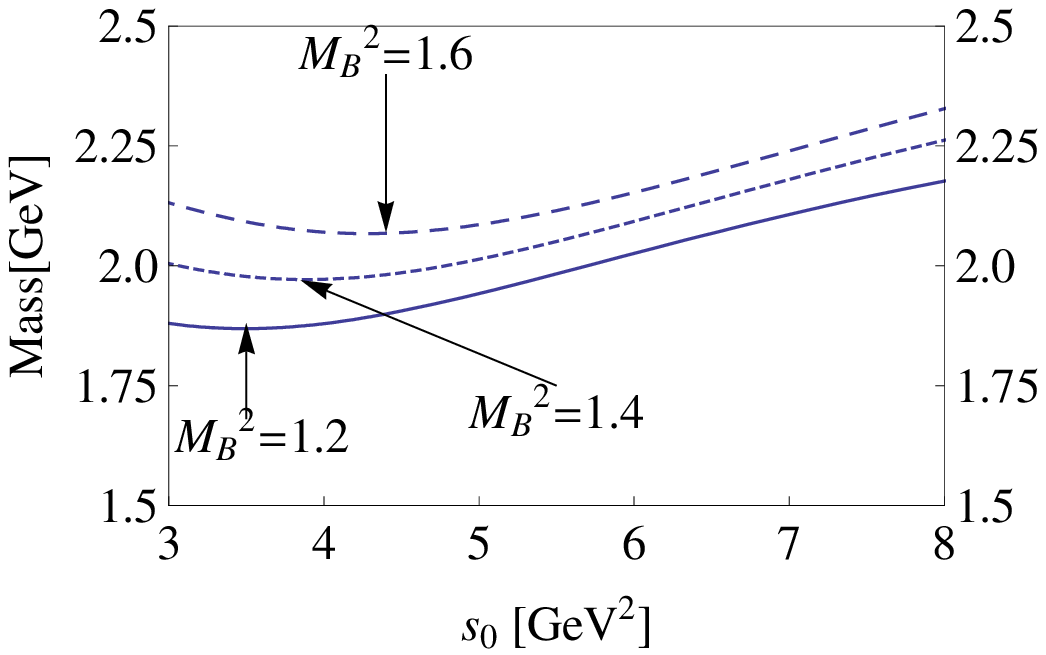}}
\end{tabular}
\caption{The variation of $M_X$ with $M^2_B$ (Left) and $s_0$
(Right) for the current $\eta_{7}$ using SVZ's gluon condensate
values.} \label{fig:eta7Shifman}
\end{center}
\end{figure}

In order to study the sensitivity of the sum rule to the
condensate values, we adopt two sets of the gluon condensate
values in our numerical analysis.  One set is from Ioffe's recent
review \cite{gluon1}: $\langle g_s^2GG \rangle=(0.20\pm0.16)
~\mbox{GeV}^4$, $\langle g_s^3 f GGG\rangle=0.12~\mbox{GeV}^6$. We
also use the original SVZ values \cite{Shifman:1978bx}: $\langle
g_s^2GG \rangle=(0.48\pm0.14) ~\mbox{GeV}^4$, $\langle g_s^3 f
GGG\rangle=0.045~\mbox{GeV}^6$. The working regions of the sum
rules with the above two sets of gluon condensates and $s_0=7$
GeV$^2$ are listed in Table \ref{tab1}.
\begin{table}
\begin{tabular}{c|c|c}
$\diagdown$& [$M^2_{\mbox{min}}$, $M^2_{\mbox{max}}$](\mbox{SVZ})
& [$M^2_{\mbox{min}}$,
$M^2_{\mbox{max}}$] (\mbox{Ioffe}) \\
\hline $\rho_{1-2}$ & $0.77\sim1.50$ & $0.90\sim1.68$\\
\hline $\rho_{3-4}$ & $1.22\sim1.90$ & $1.40\sim1.65$\\
\hline $\rho_{5-6}$ & $1.05\sim1.77$ & $1.55\sim1.74$\\
\hline $\rho_{7}$ & $1.10\sim1.85$ & $1.50\sim1.75$
\end{tabular}
\caption{The working region of $M_B^2$ with Ioffe's and SVZ's
gluon condensates and $s_0=7$ GeV$^2$. \label{tab1}}
\end{table}
The working region of the sum rule is very narrow even with
$s_0=7$ GeV$^2$. The variation of $M_X$ with $M_B^2$ and $s_0$ is
shown in Figs. \ref{fig:eta12Ioffe}-\ref{fig:eta7Ioffe} for the
interpolating currents $\eta_{1-2}$, $\eta_{3-4}$, $\eta_{5-6}$,
$\eta_{7}$ respectively using Ioffe's gluon condensate values. The
variation of $M_X$ with $M_B^2$ and $s_0$ and SVZ's gluon
condensate values is presented in
Figs.\ref{fig:eta12Shifman}-\ref{fig:eta7Shifman}.

For a genuine hadron state, one expects that the extracted mass
from the sum rule analysis is stable with the reasonable variation
of the Borel parameter and the continuum threshold. In other
words, there should exists dual stability in $M_B^2$ and $s_0$ in
the working region of $M_B^2$. From all these figures we notice
none of the mass curves satisfy the stability requirement. These
interpolating currents do not support a low-lying resonant signal.

%
%
%
%
%
%
%

%
%
%
%
%
%
%
%
%

\section{Conclusion}\label{sec:summary}

The exotic state with $J^{PC} = 0^{--}$ can not be composed of a
pair of gluons nor $q\bar q$. In order to explore the possible
existence of these interesting states, we first construct the
tetraquark type interpolating operators systematically. As a
byproduct, we notice that the $J^{PC} = 0^{+-}$ tetraquark
operators without derivatives do not exist. Then we make the
operator product expansion and extract the spectral density. The
gluon condensate becomes the dominant power correction. Usually
the four-quark type of condensates $\langle\bar{q}q\rangle^2$,
$\langle\bar{q}q\rangle$ $\langle g_s \bar{q}\sigma Gq\rangle$ and
$\langle g_s \bar{q}\sigma Gq\rangle^2$ are the dominant
nonperturbative corrections in the multiquark sum rules. However
these terms vanish because of the special Lorentz structure
imposed by the exotic $0^{--}$ quantum numbers.

Within the framework of the SVZ sum rule, we note that the absence
of the $\langle\bar{q}q\rangle^2$, $\langle\bar{q}q\rangle$
$\langle g_s \bar{q}\sigma Gq\rangle$ and $\langle g_s
\bar{q}\sigma Gq\rangle^2$ terms destabilize the sum rule. There
does not exist stability in either $M_B^2$ or $s_0$ in the working
region of $M_B^2$. Therefore we conclude that none of these
independent interpolating currents support a resonant signal below
2 GeV, which is consistent with the current experimental
measurement \cite{Amsler:2008zzb}.

\section*{Acknowledgments}

The authors are grateful to Professor Wei-Zhen Deng for useful
discussions. This project was supported by the National Natural
Science Foundation of China under Grants 10625521, 10721063 and
Ministry of Science and Technology of China (2009CB825200).


\appendix

\section{Interpolating Currents in $(\bar{q}q)(\bar{q}q)$ Basis}\label{app}

For $6_F \otimes \bar{6}_F~(S)$:
\begin{eqnarray}
 \nonumber\eta^{(S)(1)}_{m}&=&(\bar{q}_{1a}\gamma_{\mu}q_{1a})(\bar{q}_{2b}\gamma^{\mu}\gamma_5 q_{2b})
 +(\bar{q}_{1a}\gamma_{\mu}\gamma_{5}q_{1a})(\bar{q}_{2b}\gamma^{\mu}q_{2b})
 +(\bar{q}_{1a}\gamma_{\mu}q_{2a})(\bar{q}_{2b}\gamma^{\mu}\gamma_{5}q_{1b})
 +(\bar{q}_{1a}\gamma_{\mu}\gamma_{5}q_{2a})(\bar{q}_{2b}\gamma^{\mu}q_{1b}),
\\
 \nonumber\eta^{(S)(8)}_{m}&=&\lambda_{ab}\lambda_{cd}\{(\bar{q}_{1a}\gamma_{\mu}q_{1b})(\bar{q}_{2c}\gamma^{\mu}\gamma_{5}q_{2d})
 +(\bar{q}_{1a}\gamma_{\mu}\gamma_{5}q_{1b})(\bar{q}_{2d}\gamma^{\mu}q_{2d})
 +(\bar{q}_{1a}\gamma_{\mu}q_{2b})(\bar{q}_{2c}\gamma^{\mu}\gamma_{5}q_{1d})
 +(\bar{q}_{1a}\gamma_{\mu}\gamma_{5}q_{2b})(\bar{q}_{2c}\gamma^{\mu}q_{1d})\},
\end{eqnarray}
For $(\bar{3}_F \otimes \bar{6}_F) \oplus (6_F \otimes 3_F)~(M)$:
\begin{eqnarray}
\nonumber\
\eta^{(M)(1)}_{1m}&=&(\bar{q}_{1a}q_{1a})(\bar{q}_{2b}\gamma_5q_{2b})-(\bar{q}_{1a}\gamma_5q_{1a})(\bar{q}_{2b}q_{2b}),\\
\nonumber\
\eta^{(M)(8)}_{1m}&=&\lambda_{ab}\lambda_{cd}\{(\bar{q}_{1a}q_{1b})(\bar{q}_{2c}\gamma_5q_{2d})
-(\bar{q}_{1a}\gamma_5q_{1b})(\bar{q}_{2c}q_{2d})\},\\
\nonumber\
\eta^{(M)(1)}_{2m}&=&(\bar{q}_{1a}\gamma_{\mu}q_{1a})(\bar{q}_{2b}\gamma^{\mu}\gamma_5q_{2b})
-(\bar{q}_{1a}\gamma_{\mu}\gamma_5q_{1a})(\bar{q}_{2b}\gamma^{\mu}q_{2b}),\\
\nonumber\
\eta^{(M)(8)}_{2m}&=&\lambda_{ab}\lambda_{cd}\{(\bar{q}_{1a}\gamma_{\mu}q_{1b})(\bar{q}_{2c}\gamma^{\mu}\gamma_5q_{2d})
-(\bar{q}_{1a}\gamma_{\mu}\gamma_5q_{1b})(\bar{q}_{2c}\gamma^{\mu}q_{2d})\},
\end{eqnarray}
For $\bar{3}_F \otimes 3_F~(A)$:
\begin{eqnarray}
 \nonumber\eta^{(A)(1)}_{m}&=&(\bar{q}_{1a}\gamma_{\mu}q_{1a})(\bar{q}_{2b}\gamma^{\mu}\gamma_5 q_{2b})
 +(\bar{q}_{1a}\gamma_{\mu}\gamma_{5}q_{1a})(\bar{q}_{2b}\gamma^{\mu}q_{2b})
 -(\bar{q}_{1a}\gamma_{\mu}q_{2a})(\bar{q}_{2b}\gamma^{\mu}\gamma_{5}q_{1b})
 -(\bar{q}_{1a}\gamma_{\mu}\gamma_{5}q_{2a})(\bar{q}_{2b}\gamma^{\mu}q_{1b}),
\\
 \nonumber\eta^{(A)(8)}_{m}&=&\lambda_{ab}\lambda_{cd}\{(\bar{q}_{1a}\gamma_{\mu}q_{1b})(\bar{q}_{2c}\gamma^{\mu}\gamma_{5}q_{2d})
 +(\bar{q}_{1a}\gamma_{\mu}\gamma_{5}q_{1b})(\bar{q}_{2d}\gamma^{\mu}q_{2d})
 -(\bar{q}_{1a}\gamma_{\mu}q_{2b})(\bar{q}_{2c}\gamma^{\mu}\gamma_{5}q_{1d})
 -(\bar{q}_{1a}\gamma_{\mu}\gamma_{5}q_{2b})(\bar{q}_{2c}\gamma^{\mu}q_{1d})\},
\end{eqnarray}
where the indices $(1)$, $(8)$ represent the color singlet and
octet. Now we get eight mesonic currents. Then we introduce the
formula of the interchange of the color indices:
\begin{eqnarray}
\nonumber(q_{1a}q_{2b}\bar{q}_{3a}\bar{q}_{4b})&=&\frac{1}{3}(q_{1a}q_{2b}\bar{q}_{3b}\bar{q}_{4a})
+\frac{1}{2}\lambda_{ab}\lambda_{cd}(q_{1a}q_{2c}\bar{q}_{3d}\bar{q}_{4b}),\\
\lambda_{ab}\lambda_{cd}(q_{1a}q_{2c}\bar{q}_{3b}\bar{q}_{4d})&=&\frac{16}{9}(q_{1a}q_{2b}\bar{q}_{3b}\bar{q}_{4a})
-\frac{1}{3}\lambda_{ab}\lambda_{cd}(q_{1a}q_{2c}\bar{q}_{3d}\bar{q}_{4b}),
\end{eqnarray}
Next, we perform the Fierz rearrangement in the Lorrentz indices
with the formula
\begin{eqnarray}
(\bar{a}b)(\bar{b}a)=\frac{1}{4}(\bar{a}a)(\bar{b}b)+\frac{1}{4}(\bar{a}\gamma_5a)(\bar{b}\gamma_5b)
+\frac{1}{4}(\bar{a}\gamma_{\mu}a)(\bar{b}\gamma^{\mu}b)-\frac{1}{4}(\bar{a}\gamma_5\gamma_{\mu}a)(\bar{b}\gamma_5\gamma^{\mu}b)
+\frac{1}{8}(\bar{a}\sigma_{\mu\nu}a)(\bar{b}\sigma^{\mu\nu}b),
\end{eqnarray}
For example, we have
\begin{eqnarray}
\nonumber(q_{1a}^TCq_{2b})(\bar{q}_{3a}\gamma_5C\bar{q}^T_{4b})&=&-\frac{1}{4}(q_{1a}^TC\gamma_5C\bar{q}^T_{4b})(\bar{q}_{3a}q_{2b})
-\frac{1}{4}(q_{1a}^TC\gamma_{\mu}\gamma_5C\bar{q}^T_{4b})(\bar{q}_{3a}\gamma^{\mu}q_{2b})\\
\nonumber&&-\frac{1}{8}(q_{1a}^TC\sigma_{\mu\nu}\gamma_5C\bar{q}^T_{4b})(\bar{q}_{3a}\sigma^{\mu\nu}q_{2b})
+\frac{1}{4}(q_{1a}^TC\gamma_{\mu}\gamma_5\gamma_5C\bar{q}^T_{4b})(\bar{q}_{3a}\gamma^{\mu}\gamma_5q_{2b})\\
\nonumber&&-\frac{1}{4}(q_{1a}^TC\gamma_5\gamma_5C\bar{q}^T_{4b})(\bar{q}_{3a}\gamma_5q_{2b})\\
&=&-\frac{1}{4}(\bar{q}_{4b}\gamma_5q_{1a})(\bar{q}_{3a}q_{2b})-\frac{1}{4}(\bar{q}_{4b}\gamma_{\mu}\gamma_5q_{1a})(\bar{q}_{3a}\gamma^{\mu}q_{2b})\\
\nonumber&&+\frac{1}{8}(\bar{q}_{4b}\sigma_{\mu\nu}\gamma_5q_{1a})(\bar{q}_{3a}\sigma^{\mu\nu}q_{2b})-\frac{1}{4}(\bar{q}_{4b}\gamma_{\mu}q_{1a})(\bar{q}_{3a}\gamma^{\mu}\gamma_5q_{2b})\\
\nonumber&&-\frac{1}{4}(\bar{q}_{4b}q_{1a})(\bar{q}_{3a}\gamma_5q_{2b}),
\end{eqnarray}
There are only four independent currents among those eight mesonic
currents. Any four currents are independent and can be expressed
by the other four.
\begin{eqnarray}
\nonumber\eta^{(S)(8)}_{m}&=&\frac{4}{3}\eta^{(S)(1)}_{m},\\
\nonumber\eta^{(M)(8)}_{1m}&=&-\frac{2}{3}\eta^{(M)(1)}_{1m}-\eta^{(M)(1)}_{2m},\\
\nonumber\eta^{(M)(8)}_{2m}&=&-4\eta^{(M)(1)}_{1m}-\frac{2}{3}\eta^{(M)(1)}_{2m},\\
\nonumber\eta^{(A)(8)}_{m}&=&-\frac{8}{3}\eta^{(A)(1)}_{m},
\end{eqnarray}
We establish the relations between the diquark currents and the
mesonic currents using the Fierz transformation. For instance, we
can verify the relations
\begin{eqnarray}
\nonumber\eta^{(S)(1)}_{m}&=&-2\eta^S_d,\\
\nonumber\eta^{(M)(1)}_{1m}&=&\frac{1}{4}\eta^M_{1d}+\frac{1}{4}\eta^M_{2d},\\
\nonumber\eta^{(M)(1)}_{2m}&=&-\frac{1}{2}\eta^M_{1d}+\frac{1}{2}\eta^M_{2d},\\
\nonumber\eta^{(A)(1)}_{m}&=&-2\eta^A_d.
\end{eqnarray}

\section{Finite energy sum rule}\label{sec:FESR}
%

\begin{figure}[hbt]
\begin{center}
\begin{tabular}{lr}
\scalebox{0.55}{\includegraphics{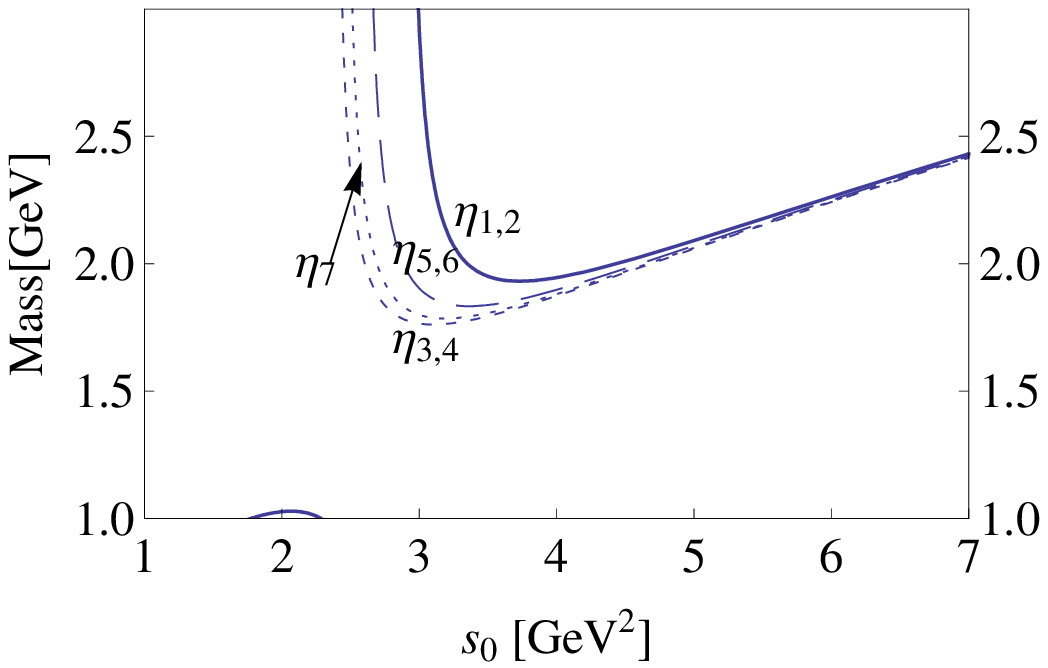}}&
\scalebox{0.55}{\includegraphics{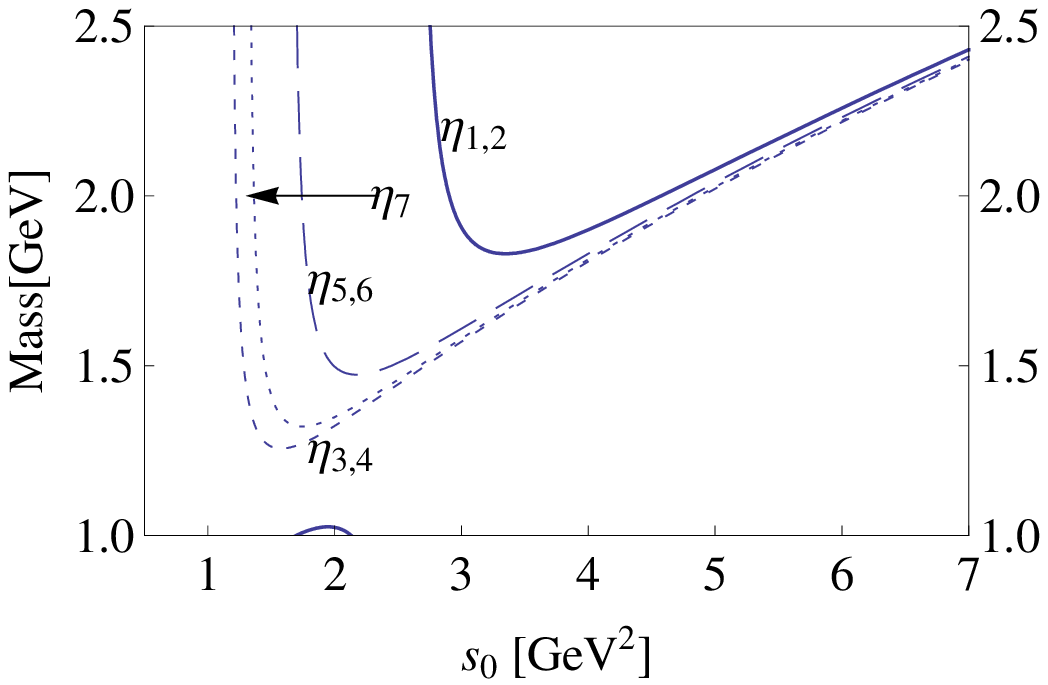}}
\end{tabular}
\caption{The variation of $M_X$ with $s_0$ and $n=0$ from the
finite energy sum rule. The left and right diagrams correspond to
Ioffe's and SVZ's gluon condensate values respectively.}
\label{fig:FESR}
\end{center}
\end{figure}

Sometimes the finite energy sum rule is also employed in the
numerical analysis. One first defines the $n$th moment using the
spectral density
\begin{eqnarray}
W(n,s_0) = \int^{s_0}_0 \rho(s)s^n ds. \label{moment}
\end{eqnarray}
With the quark-hadron duality, we have
\begin{eqnarray}
W(n,s_0)|_{Hadron}=W(n,s_0)|_{OPE}.
\end{eqnarray}
The mass of the ground state can be obtained as
\begin{eqnarray}
M_X^2(n,s_0)=\frac{W(n+1,s_0)}{W(n,s_0)}.
\end{eqnarray}
We have plotted the variation of $M_X$ with $s_0$ for all the
seven interpolating currents in Fig. \ref{fig:FESR}. The left and
right diagrams correspond to Ioffe's and SVZ's gluon condensate
values respectively. It seems that there exists a minimum of $M_X$
for each current. However, a reasonable sum rule requires that the
operator product expansion should converge well. In other words,
we require that the two-gluon power correction be less than one
third of the perturbative term and the tri-gluon power correction
less than one third of two-gluon power correction in $W(0, s_0)$,
which leads to the working window of this finite energy sum rule
as:
\begin{displaymath}
\begin{array}{c|c|c}
\diagdown& s_0(\mbox{SVZ}) & s_0(\mbox{Ioffe}) \\
\hline \rho_{1-2} & 4.0 & 7.0\\
\hline \rho_{3-4} & 4.2 & 5.7\\
\hline \rho_{5-6} & 4.0 & 7.0\\
\hline \rho_{7} & 4.9 & 6.0
\end{array}
\end{displaymath}
Clearly for each current the minimum of the mass curve lies
outside of the working region in both of the figures and is not a
real resonant signal. Starting from 4.0 GeV$^2$, each mass curve
grows monotonically with $s_0$. Thus, there does not exist a
resonant signal for every interpolating current.

%
%
%

\end{document}